\documentclass[5p]{elsarticle}

\usepackage[utf8]{inputenc}
\usepackage{amsfonts}
\usepackage{amsmath}
\usepackage{multirow}
\usepackage{caption}
\usepackage{stfloats}
\usepackage{subfig}
\usepackage[hidelinks]{hyperref}
\usepackage{xcolor}
\usepackage{graphicx}

\journal{a journal to be decided}
\begin{document}
\begin{frontmatter}
\title{MBD: Multi b-value Denoising of Diffusion Magnetic Resonance Images}%

\author{Jakub Jurek$^1$}
\ead{jakub.jurek@p.lodz.pl}
\author{Andrzej Materka$^1$}
\author{Kamil Ludwisiak$^2$}
\author{Agata Majos$^3$}
\author{Filip Szczepankiewicz$^4$}

\address{$^1$Institute of Electronics, Lodz University of Technology, Aleja Politechniki 10, PL-93590 Lodz, Poland}
\address{$^2$Department of Diagnostic Imaging, Independent Public Health Care, Central Clinical Hospital, Medical University of Lodz, Pomorska 251, PL-92213 Lodz, Poland}
\address{$^3$Department of Radiology, Medical University of Lodz, Lodz, Poland}
\address{$^4$Medical Radiation Physics, Lund University, Barngatan 4, 22185 Lund, Sweden}

\begin{abstract}

We propose a novel approach to denoising diffusion magnetic resonance images (dMRI) using convolutional neural networks, that exploits the benefits of data acquired at multiple b-values to offset the need for many redundant observations. Denoising is especially relevant in dMRI since noise can have a deleterious impact on both quantification accuracy and image preprocessing. The most successful methods proposed to date, like Marchenko-Pastur Principal Component Analysis (MPPCA) denoising, are tailored to diffusion-weighting repeated for many encoding directions. They exploit high redundancy of the dataset that oversamples the diffusion-encoding direction space, since many directions have collinear components. 

However, there are many dMRI techniques that do not entail a large number of encoding directions or repetitions, and are therefore less suited to this approach. For example, clinical dMRI exams may include as few as three encoding directions, with low or negligible data redundancy across directions. Moreover, promising new dMRI approaches, like spherical b-tensor encoding (STE), benefit from high b-values while sensitizing the signal to diffusion along all directions in just a single shot. 

We introduce a convolutional neural network approach that we call multi-b-value-based denoising (MBD). MBD exploits the similarity in diffusion-weighted images (DWI) across different b-values but along the same diffusion encoding direction. It allows denoising of diffusion images with high noise variance while avoiding blurring, and using just a small number input images.

We compare MBD to other convolutional neural network approaches and MPPCA on a synthetic dataset, featuring a brain with and without lesions, and two brain imaging datasets of healthy volunteers: a clinical scan at 1.5 Tesla and the maximum b-value of 1000 s/mm$^2$ and an STE scan at 3 Tesla with the maximum b-value of 4000 s/mm$^2$. We show that MBD, using just a few input images, outperforms reference methods and can reliably estimate the intensity of dMRI images even for considerably noisy images.

\textbf{Declarations of interest}: None.
\end{abstract}


\end{frontmatter}

\section{Introduction}

\subsection{Background}
Diffusion magnetic resonance imaging (dMRI) is a unique imaging tool for probing microstructural properties of tissues, occurring at scales much smaller than the actual spatial resolution of the images~\cite{Jelescu2020}. Despite the interest of the clinical diagnostics in this modality, many dMRI preprocessing tasks have not been solved yet~\cite{Tax2022}. A persisting problem is noise which adversely affects both qualitative (i.e. visual) and quantitative (accuracy and precision) dMRI analysis. 

In MRI, noise primarily arises from inherent thermal oscillations of electrons in tissues, that additively combine with the useful nuclear magnetic resonance signals used to reconstruct images. The use of echo-planar imaging with short acquisition times and the attenuative nature of diffusion-weighting concept results in a relatively low amplitude of dMRI signals compared to other MRI techniques. dMRI images and their analysis is thus particularly sensitive to noise.

To recover even the most basic diffusion-based quantities, the measurement must employ multiple diffusion-weighting strengths. This strength is described by a factor called the b-value. The b-value of 0 s/mm$^2$ means no diffusion-weighting and no corresponding contrast. Increasing the b-value produces more diffusional contrast, but decreases the amplitude of signals. Therefore, data acquired at ever higher b-values will suffer from lower signal-to-noise ratio (SNR). 

Signal precision and accuracy noise variance and mean value, respectively. In MRI, the mean value is non-zero for magnitude images~\cite{Gudbjartsson1995} and is referred to as the Rician bias or noise floor. Parameters based on the magnitude signal feature a systematic error. In radiological interpretation, Rician bias results in a descreased contrast in low-signal areas. High variance relates to the uncertainty of the signal value from a single measurement and it yields a grainy appearance if images and parameter maps. Issues with precision can be parried by repeating the measurement several times, however, poor accuracy does not have a similar remedy. Keeping dMRI resolution low compared to other MRI techniques improves precision and accuracy, but may impair the analysis of small structures.

\subsection{dMRI denoising}

Denoising is a post-processing procedure that can improve the precision and accuracy of dMRI measurements. It can be applied in the complex or magnitude domain. Complex denoising can both improve precision and reduce Rician bias. However, since complex denoising imposes additional challenges such as phase correction~\cite{CorderoGrande2019}, we focus more on magnitude-domain denoising. We aim to improve precision without increasing the bias. 

Multiple approaches to dMRI denoising, both general-purpose and specialized, were proposed~\cite{ManzanoPatron2024}. General-purpose algorithms are not tailored to dMRI and include methods such as non-local means~\cite{Buades2005, Coupe2008}, total variation~\cite{Liu2014} and convolutional-neural-network-based denoising~\cite{Muckley2020}. In our previous studies, we developed methods for dMRI denoising using convolutional neural networks (CNNs). Complex-domain denoising was considered, using transfer learning on synthetic data~\cite{Jurek2023} and Noise2Noise~\cite{Lehtinen2018} (N2N) after phase correction~\cite{Jurek2023a}. Issues were observed related to blurring and phase correction errors. 

By contrast, specialized algorithms are suited to specific dMRI dataset properties. In particular, they often make use of the redundancy of datasets where multiple diffusion-encoding directions were sampled, for example diffusion tensor or diffusion kurtosis imaging (DTI, DKI). Examples of those algorithms are MPPCA~\cite{Veraart2016} with its variation tMPPCA~\cite{Olesen2023}, Patch2Self~\cite{Fadnavis2020} and SDnDTI~\cite{Tian2022}. 

MPPCA (Marchenko-Pastur Principal Component Analysis) is based on identifying and nullifying pure-noise principle components in small image patches~\cite{Veraart2016}. SNR improvement in MPPCA in theory is at most the improvement of averaging, but MPPCA does not require scan repetitions.

Patch2Self is based on linear regression~\cite{Fadnavis2020}. It uses all-but-one images with different diffusion-encoding directions to predict the one held out. This poses a risk of losing details that are well represented in just a few spatial-encoding directions. A regressor is trained for each image patch, diffusion direction and b-value. Training needs to be repeated for each patient. 

SDnDTI is based on a convolutional neural network (CNN)~\cite{Tian2022}. It creates subsets of a multidirectional dataset and uses interpolation, for each subset, to predict images at diffusion-encoding directions from other subsets. These interpolated repetitions have independent noise and are averaged to obtain a less noisy reference for training a denoiser CNN. The network is fed with a b=0s/mm$^2$ image and a set of images at one higher b-value, with different diffusion-encoding directions. 

Despite the optimistic results of the described redundancy-based methods, it may be challenging to obtain satisfactory denoising quality for data with low redundacy, e.g. when scan repetitions or diffusion-encoding directions are not involved. This is the case in standard clinical dMRI, which often involves just a small number of repetitions and diffusion-encoding directions, since the goal is to rapidly produce parameter maps such as the apparent diffusion coefficient (ADC) map. Moreover, the employed directions are nearly orthogonal which causes minimal redundancy, especially in tissue with high diffusion anisotropy, like white matter. Also some more advanced, modern dMRI variations do not involve multidirectional diffusion encoding. An example is diffusion imaging with spherical b-tensor encoding (STE)~\cite{Westin2016, Tax2020}. It does not require multidirectional acquisition as it sensitizes image contrast to all diffusion directions at the same time and provides complementary contrast at high b-values. However, STE features low SNR.

From the state of the art it follows that new and fast denoising methods are required for clinical dMRI and for some modern variations like STE. Convolutional neural networks, once trained, are fast in inference, but so far they either required redundancy, availability of multiple diffusion-encoding directions (SDnDTI~\cite{Tian2022}) or had blurring issues~\cite{Jurek2023}. In this paper, we propose a method which integrates multiple concepts. We use a CNN as a denoiser, trained with the Noise2Noise approach that does not require noiseless training targets~\cite{Lehtinen2018}. We feed the network with images acquired at multiple b-values. This approach is known as guided denoising~\cite{Kaye2020, Mohammadi2023, Pfaff2024}. It was also was partly featured in~\cite{Tian2022}, but remained unrecognized and unexplored. Despite the concept was already applied, it was not properly investigated. It also lacks quantitative evaluation, especially for very high b-values and possibly anisotropic tissues. We deepen the understanding and expand the application of guided denoising in dMRI. Specifically, we feed the neural network with a b0 (we denote the images acquired at particular b-value of X s/mm$^2$ as bX, e.g. b0, b4000 etc.), an intermediate-b-value and a high-b-value image. We demonstrate that this gives superior results to ordinary training with a single input an two inputs. We propose to call this variation of guided denoising multi-b-value-based denoising (MBD). We conducted multiple experiments on magnitude brain images, both synthetic and real. MBD is compared to other learning-based methods: ordinary Noise2Noise and CNN-based extrapolation, to algebraic extrapolation and to MPPCA. 

\section{Methods and Materials}

\subsection{Noise to Noise and multi-b-value-based denoising}

\begin{figure*}[ht!]
\centering
\subfloat[]{\includegraphics[width=\textwidth]{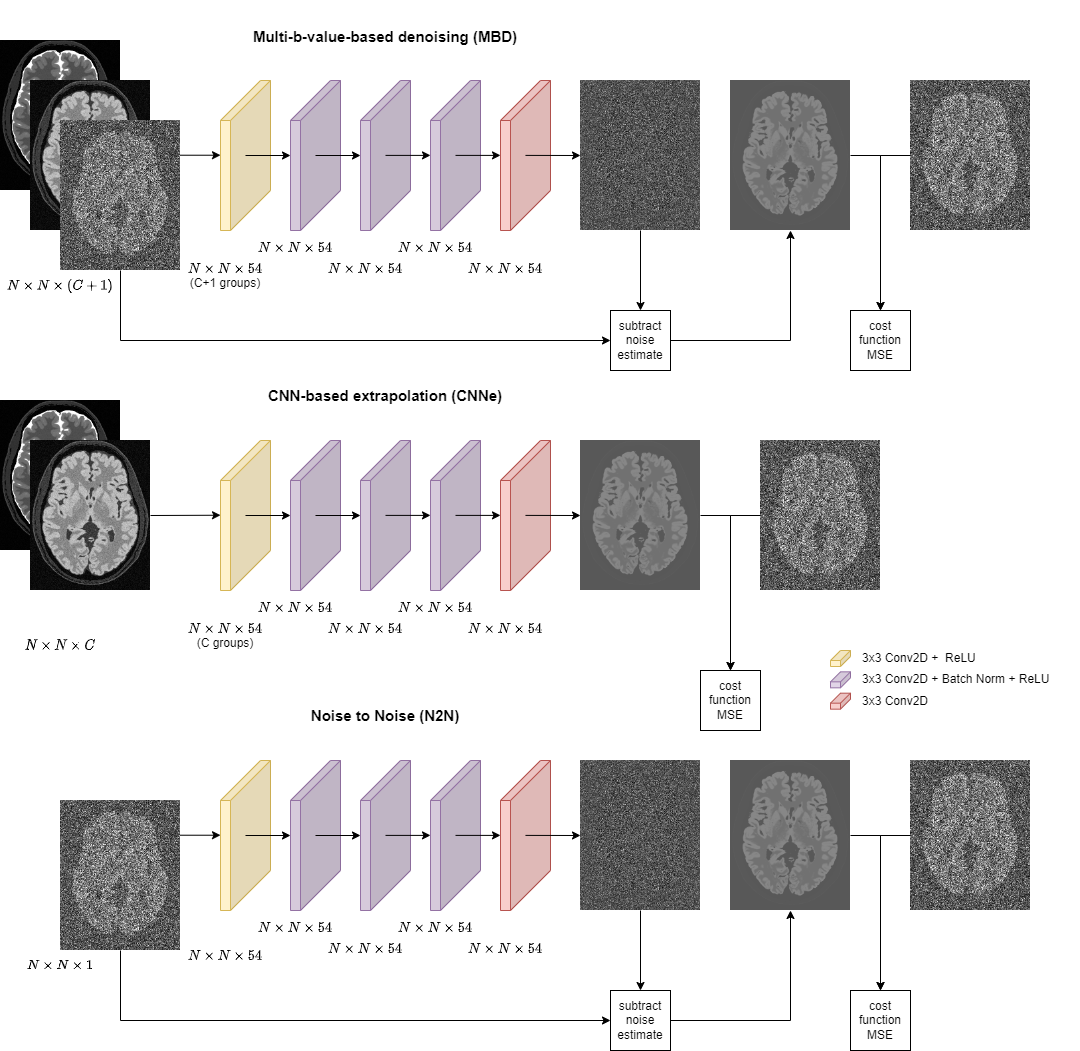}}\hfill
\caption{Configuration and architecture of the MBD, CNNe and N2N convolutional neural networks. Note the differences in the input, while the architecture and target data are the same. $C$ is the number of input images (channels) for CNNe. $C=2$ case is presented.}
\label{architectures}
\end{figure*}

Consider $\hat{x_i}$ to be noisy images and $y_i$ to be noiseless training targets. In classical learning-based denoising, the training objective is to find model parameters that minimize the loss function $L$ over the training set:
\begin{equation}
\underset{\theta}{\textup{argmin}}\sum_i L(f_{\theta}(\hat{x_i},y_i))
\end{equation}
A popular choice for $L$ is the mean squared error (MSE), denotes as $L_2$, which has the property of estimating the population mean. If that is the case, substituting noiseless targets $y_i$ with their noisy versions $\hat{y_i}$, assuming a large training dataset, should not change the optimum set of parameters $\theta$:
\begin{equation}
\underset{\theta}{\textup{argmin}}\sum_i L(f_{\theta}(\hat{x_i},\hat{y_i})) = \underset{\theta}{\textup{argmin}}\sum_i L(f_{\theta}(\hat{x_i},y_i)).
\end{equation}
This shows that noisy images can be used as targets to train a denoiser and is the basis for Noise2Noise~\cite{Lehtinen2018}. For images corrupted with zero-mean Gaussian noise, such as complex-valued MRI, one can expect no bias from N2N denoising. For Rician noise, however, N2N will reduce noise variance, but will keep the Rician bias since it is the expected intensity value of a voxel. A neural network architecture using N2N is shown in the bottom row in Figure~\ref{architectures}. Typically, as an individual training vector, the network is fed with a single image, and a repeated measurement of that image, featuring independent noise, is set as the learning target.

In MBD, we modify the input to the N2N network~\cite{Kaye2020} and supply a sequence of b-values as individual input channels to provide more information about the structure and diffusional properties of the tissues (top row in Figure~\ref{architectures}). There is no theoretical limit on the number of the input channels, however, we currently investigated cases that feature at most three b-values, corresponding to the practice of one of our partner hospitals. For example, denoising of b4000 images is learned based on pairs involving b0, b1000 and b4000 images as input and the second repetition of b4000 images as targets. The choice of optimal b-values was not investigated in this paper (see Discussion), but we demonstrate using different datasets that the proposed method performed better than the reference methods regardless of the b-values involved.

\subsection{Extrapolation methods}

Assuming some model of dMRI signal intensity, images for an arbitrary b-value can be obtained using extrapolation. This method is offered as a post-processing feature in some commercial MRI machines and is used by radiologists in prostate cancer imaging, for instance~\cite{Bittencourt2014}. To prove that MBD is superior to trivial extrapolation, we implemented two reference extrapolation methods. 

The classical algebraic extrapolation (ALGe) is based on dMRI signal models. It was considered in the two-point version, i.e. the image at the desired b-value is extrapolated from two images at lower b-values. Using Eq.~\ref{see}, $S(v, b)$ for any $b$ can be calculated through estimated diffusivity as:
\begin{eqnarray}
D \approx \frac{\textup{log}(S_N(v, b_1))-\textup{log}(S_N(v, b_2))}{b_2-b_1}\\
S_N(v, b) \approx S_N(v, b_1)\cdot e^{-(b-b_1)\cdot D}
\label{e:alge}
\end{eqnarray}

For a better comparison with MBD, we propose a CNN-based extrapolation method (CNNe), similar in form to N2N and MBD. Referring to Eq.~\ref{e:alge}, CNNe is fed with image at $b_1$ and $b_2$, while the training target is an image at some other $b>b_1, b_2$ . This is shown in the middle row in Figure~\ref{architectures}. In this setting, the network is forced to 'extrapolate'. In contrast to algebraic extrapolation, CNNe is expected to have denoising properties because the target image inherently features independent noise. 

\subsection{Proposed neural network architectures for N2N, CNNe and MBD}

For all neural networks in this paper, we relied on the DnCNN architecture~\cite{Zhang2017}. DnCNN features an input 2D convolutional layer with ReLU activation and an output linear convolutional layer. Between the input and output layers, there is a user-defined number of blocks, consisting of a convolution layer, batch normalization and ReLU. In our implementation, we used three such blocks, so the total depth is five layers. All layers compute 54 features except for the final. The first layer computes features in equally-sized groups, separately for each input channel, but keeping 54 features in total. Convolutions use $3\times3$ kernels. The input is multichannel, with a different number of channels for N2N, MBD and CNNe. Originally, DnCNN estimates the noise instance and subtracts it from the input. This approach was used for MBD and N2N. For MBD, noise is estimated for the image at the desired b-value and subtracted from the corresponding input channel. CNNe directly estimates the noiseless image at the desired b-value. 

\subsection{Quantitative comparison of denoised images}

N2N provides a means of estimating the error with respect to the ground truth, even though a noiseless image is unknown. Consider a pair of noisy image repetitions. The difference of those images yields, assuming no motion or other artefacts, a noise map whose variance is twice the variance of noise in the individual images:
\begin{equation} 
\mathrm{Var}[I_1 - I_2] =  \mathrm{Var}[N_{I_1}]+\mathrm{Var}[N_{I_2}]
\end{equation} 
Variance of the image repetitions difference is the MSE value before training starts, $\mathrm{Var}[I_1 - I_2] = \mathrm{MSE}(\hat{I_2}, I_2)$. The theoretical minimum is achieved when the noise variance in the network output is 0. Hence, the minimum loss achievable in N2N is $\mathrm{MSE} = \mathrm{Var}[N_{I_2}] = \frac{\mathrm{Var}[N_{I_1}]+\mathrm{Var}[N_{I_2}]}{2}$, if both variances are equal, which can be assumed for repeated acquisition of the same slice. Alternatively, we can compute the mean absolute error (MAE), in which case the minimum is the standard deviation of the noise, $\mathrm{MAE} = \sigma[N_{I_2}]$. We use this property when comparing methods on datasets without a true reference, but with multiple noisy repetitions. 

\subsection{Clinical-like brain dMRI by linear tensor encoding}
Ten volunteers underwent a dMRI brain scan on a MAGNETOM Avanto 1.5T (Siemens Healthineers, Germany) system at the Central Teaching Hospital, Medical University of Lodz, Poland. Approval of the procedure was obtained at the Medical University of Lodz and all volunteers gave their consent to participate. Imaging parameters were: TR/TE = 5700/114 ms, FOV = 230$\times$230 mm$^2$, matrix = 160$\times$160, bandwidth = 1250 Hz/px, echo spacing = 0.89 ms, EPI factor = 160, Partial Fourier = 6/8 (phase-encoding direction). Parallel imaging (Siemens iPAT) was not used. Any filters available by the scanner user interface were turned off to ensure minimal interference in the noise~\cite{Jurek2023}. 

\begin{figure}[ht!]
\centering
\subfloat[b = 0 s/mm$^2$]{\includegraphics[width=0.32\columnwidth]{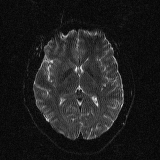}}\hfill
\subfloat[b = 300 s/mm$^2$]{\includegraphics[width=0.32\columnwidth]{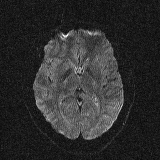}}\hfill
\subfloat[b = 1000 s/mm$^2$]{\includegraphics[width=0.32\columnwidth]{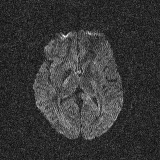}}\hfill

\caption{Example axial slice from the clinical brain dMRI, involving b-values of 0, 300 and 1000 s/mm$^2$.}
\label{f:clinicaldata}
\end{figure}

Three orthogonal (x, y, z) diffusion directions were sampled at b-values of 0, 300 and 1000 s/mm$^2$. The images have spatial resolution of 1.4$\times$1.4$\times$5 mm$^3$. For the sake of denoising quality evaluation, the number of scan repetitions was set to the maximum possible value of 32. The images were exported as raw data and reconstructed using the algorithm described in~\cite[Section 3.1]{Jurek2023}. An example slice at all b-values is shown in Figure~\ref{f:clinicaldata}.

\begin{figure}[ht!]
\centering
\subfloat[b = 0 s/mm$^2$]{\includegraphics[width=0.32\columnwidth]{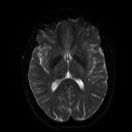}}\hfill
\subfloat[b = 1000 s/mm$^2$]{\includegraphics[width=0.32\columnwidth]{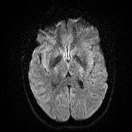}}\hfill
\subfloat[b = 4000 s/mm$^2$]{\includegraphics[width=0.32\columnwidth]{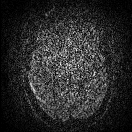}}\hfill
\caption{Example axial slice from the STE brain dMRI, involving b-values of 0, 1000 and 4000 s/mm$^2$.}
\label{f:stedata}
\end{figure}

\subsection{Brain dMRI by spherical b-tensor encoding}
We reused the dataset described in~\cite{Vis2021}. Images were acquired using the MAGNETOM Prisma 3T (Siemens Healthineers, Germany) system. A research pulse sequence was used~\cite{Szczepankiewicz2019}. One volunteer was enrolled. However, as the original paper was targeted at dMRI super-resolution reconstruction, eight volumes were acquired with differently rotated fields of view. Namely, the FOV was rotated about the anteroposterior axis, starting in the axial orientation, and then at angles of 22.5, 45, 67.5, 90 (sagittal), 112.5, 135, 157.5 degrees. b-values of 0, 1000 and 4000 s/mm$^2$ were used with 2, 4 and 15 repetitions, respectively, for each field of view. The images have 1.58$\times$1.58$\times$7.2 mm$^3$ voxels and matrix size of 132$\times$132$\times$26. An example slice is shown in Figure~\ref{f:stedata}.

\subsection{Synthetic dMRI dataset and simulation methodology}

The Illinois Institute of Technology brain phantom (nitrc.org/project/iit) was used to simulate a DTI dataset. The brain represented as fuzzy digital maps of cerebrospinal fluid (CSF), grey matter (GM) and white matter (WM). The tissue maps define the proportional content of each tissue per voxel, in a 182$\times$218$\times$182 grid at 1$\times$1$\times$1 mm$^3$ spatial resolution. The phantom also includes a voxelwise map of the diffusion tensor. 

The spin-echo signal equation was used together with a monoexponential signal decay model to generate the intensity $S$ for every voxel $v\in V$, the set of all voxels:
\begin{equation}
S(v, b) = \sum_{t \in T} F_t(v) \cdot k\cdot \rho \cdot \mathrm{e}^{\frac{-TE}{T_2(v)}}\cdot(1-\mathrm{e}^{\frac{-TR}{T_1(v)}})\cdot \mathrm{e}^{-b\cdot D(v)},
\label{see}
\end{equation}

\noindent where $F_t(v)$ defines the proportion of tissue $t$ in voxel $v$, $k$ is the signal gain, $\rho$ is the proton density, $TE$ and $TR$ are the echo and repetition times, respectively, $T_1$ is the longitudinal relaxation time, $T_2$ is the transverse relaxation time, $b$ is the diffusion-weighting b-value, and $D$ is the apparent diffusion coefficient. For all simulations, imaging parameter values were set to $k=1000$, $TR=6700$ ms, $TE=100$ ms. $\rho$ wrt. CSF, $T_1$ {[}ms{]}, $T_2$ {[}ms{]} parameters were set to 1/0.86/0.77, 2569/833/500, 329/83/70 for the CSF/GM/WM tissues~\cite{AubertBroche2006b}. 

To obtain diffusion-weighting dependent on the chosen diffusion direction, we calculated directional diffusivity maps $D_k$ according to the formula:
\begin{equation}
D_k = \mathbf{g}^\top_k\mathbf{D}\mathbf{g}_k,
\end{equation}
with $\mathbf{g}_k$ being a unit vector in direction $k$ and $\mathbf{D}$ the diffusion tensor~\cite{Basser1994}. $D_k$ was then substituted as $D$ in Equation~\ref{see}.  

Diffusion directions were chosen using DIPY's \textit{generate\_bvecs}~\cite{Garyfallidis2014}. For a given number of directions, the function generates a random set of spatial directions, optimally distributed within a hemisphere using the electrostatic repulsion model~\cite{Bak1997, Jones2004}. A sets of 16 directions was generated. 

Noise was added to the simulated volumes using the following relation:
\begin{equation}
S_N(v, b, \sigma) =\left |  \frac{S(v, b)}{\sqrt{2}}+n_1(\sigma) + (\frac{S(v, b)}{\sqrt{2}} + n_2(\sigma))i\right |,
\end{equation}
with $n_1, n_2\sim \mathcal{N}(\mu = 0, \sigma)$ being independent random noise values, sampled from the normal distribution. Since $S_N$ is calculated as the magnitude of a complex intensity, the simulated images feature Rician noise~\cite{Gudbjartsson1995}. 

The IIT phantom does not involve any lesion regions. To evaluate how the proposed method treats small regions with different properties than their surroundings, we simulated brain lesions and integrated them with the phantom. Lesion masks and the corresponding MRI (T1w, T2w, FLAIR) were extracted from a publicly available multiple-sclerosis (MS) dataset~\cite{Muslim2022}. The method of lesion extraction is detailed in the Supplement (Figures~S1-S4).

\begin{figure}[ht!]
\centering
\subfloat[b = 0 s/mm$^2$]{\includegraphics[width=0.32\columnwidth]{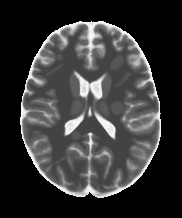}}\hfill
\subfloat[b = 1000 s/mm$^2$]{\includegraphics[width=0.32\columnwidth]{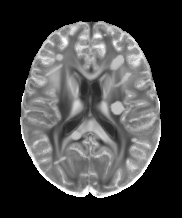}}\hfill
\subfloat[b = 4000 s/mm$^2$]{\includegraphics[width=0.32\columnwidth]{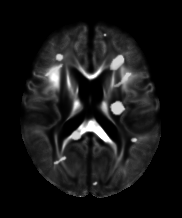}}\hfill

\subfloat[b = 0 s/mm$^2$]{\includegraphics[width=0.32\columnwidth]{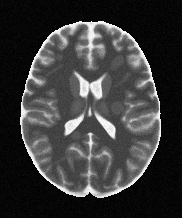}}\hfill
\subfloat[b = 1000 s/mm$^2$]{\includegraphics[width=0.32\columnwidth]{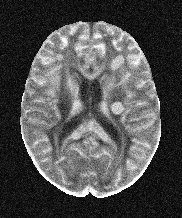}}\hfill
\subfloat[b = 4000 s/mm$^2$]{\includegraphics[width=0.32\columnwidth]{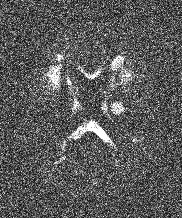}}\hfill

\caption{Example axial slice from synthetic brain DTI. Noiseless reference (top) and noisy images with $\sigma= 7\%$ of the true white matter intensity at b=0 (bottom) are shown. Several differently shaped lesions are visible, with $f=0.75$, $D_1=400\frac{mm^2}{s}$, $D_2=600 \frac{mm^2}{s}$, $\Delta T_2=15$ ms. They are hyperintense at all b-values compared to normal white matter. Exponential SNR drop is well visible for the b4000 image.}
\label{f:dtidata}
\end{figure}

Diffusional parameters of lesions were randomly selected from predefined intervals. We assumed lesions follow the bi-exponential dMRI signal model: 
\begin{equation}
S(b) = k\cdot \rho \cdot e^{\frac{-TE}{T_2}}\cdot(1-e^{\frac{-TR}{T_1}})\cdot (f\cdot e^{-b\cdot D_1}+(1-f)\cdot e^{-b\cdot D_2})
\label{biexp}
\end{equation}
where $f\in [0, 1]$ and $D_1$ and $D_2$ are diffusion coefficients of different tissue compartments within a voxel.

Since the IIT phantom includes one brain volume only, the training set was composed of 8 instances of this volume, differing in the diffusion direction. A different direction was used for the test set. Additionally, to enable fair testing, training lesions were chosen from a different lesion subset than test lesions.

Each lesion in the training set was assigned random tissue parameters, namely the diffusivities $D_1$ and $D_2$, the weighting factor $f$ and an additive constant $\Delta T_2$, modifying the signal equation for lesion tissue to:
\begin{equation}
S^{lesion}(b) = k\cdot \rho \cdot e^{\frac{-TE}{T_2+\Delta T_2}}\cdot(1-e^{\frac{-TR}{T_1}})\cdot (f\cdot e^{-b\cdot D_1}+(1-f)\cdot e^{-b\cdot D_2})
\label{biexples}
\end{equation}
Parameters $\rho$, $T_1$ and $T_2$ for lesions were equal to white matter values. $D_1$ and $D_2$ were sampled uniformly from the interval [0.3,1.35), $f$ from the interval [0,1] and $\Delta T_2$ from the interval [-30,30] ms with a step of 5 ms. The latter was used to incorporate lesions visible in b0 images, both hyper- and hypointense. A slice with hyperintense lesions in a noiseless and noisy case is shown in Figure~\ref{f:dtidata}.

\subsection{Experimental settings}

\subsubsection{Denoising of in-vivo data}

MBD, CNNe and N2N networks were trained for denoising the b1000 images in the clinical dataset and the b4000 images in the STE dataset. For the clinical dataset, out of the ten patients, one was held out for testing and the remaining nine were used for training (patch size of 32$\times$32) and validation. For the STE dataset, one of the eight rotated FOV volumes was left out for testing and the remaining 7 were used for training (patch size of 33$\times$33) and validation. 

ALGe was used to extrapolate b1000 images from b0 and b300 ones for the clinical dataset and, for the STE dataset, b4000 images from b0 and b1000. 

MPPCA (DIPY, ver. 1.7.0., dipy.denoise.localpca.mppca~\cite{Garyfallidis2014}, $patch\_radius$ = 2, other parameters set as default) was applied on 5$\times$5$\times$5 image patches with b0, b300 and b1000 diffusion-weighted images (DWI; b0, b1000 and b4000 for the STE) as the features dimension.

Loss on the validation set was monitored during training to compare learning-based models. Visual evaluation for a chosen mid-brain slice was then used to compare all denoising methods. Comparison concerned denoised NEX=1 images as well as the average of the denoised images (32 repetitions for the clinical and 15 for the STE datasets). Absolute error residual maps were computed for each of the repetitions and then averaged, to reveal systematic errors. All averaging operations were preceded by rigid registration for the clinical dataset. For the STE dataset, we verified that motion was not significant and registration was not performed.

As a quantitative test, we computed the N2N mean squared error for NEX=1 images and a reference image.

\subsubsection{Configurations of MBD}

In the described in-vivo experiments, we set up MBD for denoising of the images with highest b-value in a dataset. Even though this might be useful for radiological analysis, a more stringent set of requirements is placed on the result if it is to be used for quantification, and it would benefit from denoising also the images at lower b-values. This requires images at all b-values to be denoised. 

To investigate if this approach is useful for denoising of images at intermediate or lowest b-value in a sequence of three different b-values, we trained MBD networks with different input-output settings. For a particular b-value at the network output, there are four possible input combinations. For example, for denoising of b300 images in the clinical dataset, we considered four input options: a) b300, b) b0 and b300, c) b1000 and b300, d) b0, b300 and b1000. Such a set of denoisers was trained for each b-value denoising for both the clinical and STE in-vivo datasets. Training and validation set hyperparameters were identical as in previously described experiment on these datasets.

As a means of evaluation, we used the MSE cost function computed over the validation set. Due to randomness in neural network training (resulting from random weight initialization, random data shuffling and non-deterministic algorithms within the GPU's CUDA library~\cite{Aakesson2024}), we performed each training session 10-fold and averaged the loss curves. Since the ten training sessions took different numbers of epochs in general, the average curve was cropped to the length of the shortest session.

\subsubsection{Denoising synthetic dMRI with lesions}

IIT volumes for 8 randomly chosen diffusion-encoding directions were used for learning. For each diffusion-encoding direction, lesion locations within white matter and their diffusional parameters were selected randomly. Each volume was cropped to the range of slices containing lesions, yielding 60 slices per volume. 20 random slices were held out as the validation set, and the remaining slices were used to extract non-overlapping 30$\times$30 patches that formed the set used to train the MBD, N2N and CNNe networks. For initial evaluation, we compared validation MSE loss curves obtained for each neural network.

Testing was performed on a b4000 slice simulated for a diffusion-encoding direction unused in training and validation, with different lesion shapes and positions. The chosen slice is located mid-brain and contains ten lesions, each with the same parameter set. A thousand noisy instances of the test slice were simulated, each time with different lesion parameters, but in the same location. Each instance was denoised using MBD, CNNe, N2N, MPPCA and ALGe and the absolute error was taken versus the noiseless ground truth. Error maps were averaged for the thousand instances. 

The distribution of errors for voxels with 100\% lesion fraction was studied using histograms. Mean error histogram was considered with a bin size of 0.01 and absolute error histogram with a bin size of 1.

The link between error value and lesion parameters, reflected by the intensities in the considered b0, b1000 and b4000 images, was studied using 3D scatter plots with colour-coded error values. The percentage of lesion parameter sets for which each individual method achieved best results was computed, and plotted in a 3D scatter plot to show dependence on image intensities. 

MBD, CNNe, N2N networks, MPPCA and ALGe were also used to denoise a lesion-free version of the test slice. Difference maps between the denoised lesion and lesion-free slice where computed to observe how the methods reacted to the occurrence of the lesions. Additionally, for chosen lesion parameter sets and selected region of interest from the slice, the difference maps were compared by taking the difference for MBD-CNNe and MBD-N2N pairs, which revealed areas where MBD was better than other methods.

\section{Results}

\subsection{Clinical brain dMRI denoising}

\begin{figure*}[ht!]
\centering
\subfloat{\includegraphics[width=\textwidth]{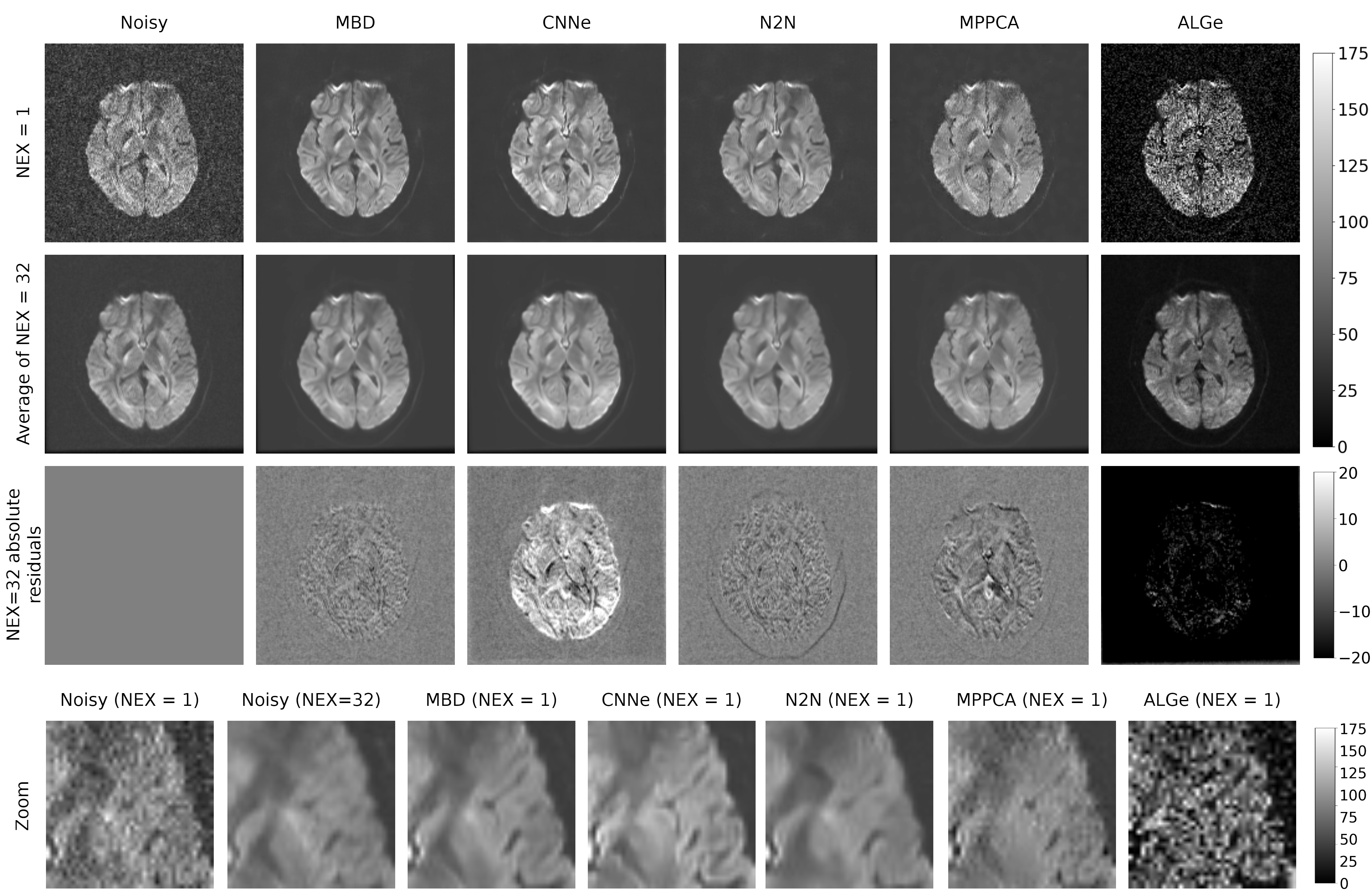}}\hfill
\caption{Comparison of MBD, N2N, CNNe, MPPCA and ALGe denoising of b1000 clinical dMRI images. Rows, from top: NEX = 1 images, NEX = 32 images, NEX=32 images (zoomed view), NEX = 32 absolute residuals. MBD shows accurate denoising and edge-preservation, even for NEX=1 images.}
\label{ckddenoised}
\end{figure*}

We compared denoised images and noise residuals across the methods. ALGe resulted in images of a more noisy appearance then the actual acquisition (Figure~\ref{ckddenoised}). General appearance of the denoised images is similar for MBD, CNNe and N2N. However, brain details are reconstructed with fine differences. A magnified patch of the brain is shown in Figure~\ref{ckddenoised} for the noisy image, average of noisy repetitions, and MBD, N2N, CNNe and MPPCA. MPPCA reduced noise variance to a certain extent, but the subjective noisiness is still visible for the single denoised image. Details of the cerebral cortex are sharpest for CNNe. This method however shows elevated signal compared to the mean image, in contrast to MBD and N2N. MBD has superior resolution to N2N as it shows more anatomical details. Despite a moderate noise level in the analysed unprocessed b1000 image, N2N blurs cerebral sulci and the visible part of globus pallidus. In comparison to the mean of noisy images, MBD has similar noise reduction but keeps the details, which is best visible for the cerebral sulci.

This is further confirmed by the averaged residual maps. CNNe results in a highly biased image. MPPCA is much better than CNNe, but it shows areas of elevated error, too. Residual maps for MBD and N2N are similar, but more structure of the image is visible in N2N residuals. The tissues around the skull, for example, which have a very low SNR, are retained for MBD while they are removed by N2N.

Finally, these observations are supported by the quantitative test performed for the unaveraged image. With 12.96 MAE value for the noisy image in reference to another noisy repetition, error values for MBD, CNNe, N2N, MPPCA and ALGe were 8.32, 9.23, 8.44, 8.78 and 33.33, respectively. 

Supplement Figure~S5 shows validation loss curves obtained during training of MBD, CNNe and N2N networks on the clinical brain dataset. MBD reached the lowest error, which proves its superiority. However, this is not perfect denoising, which is visible as the distance of the MBD loss curve from the theoretical minimum.


\subsection{Denoising of STE brain images}\label{s:stedeno}

\begin{figure*}[ht!]
\centering
\subfloat{\includegraphics[width=\textwidth]{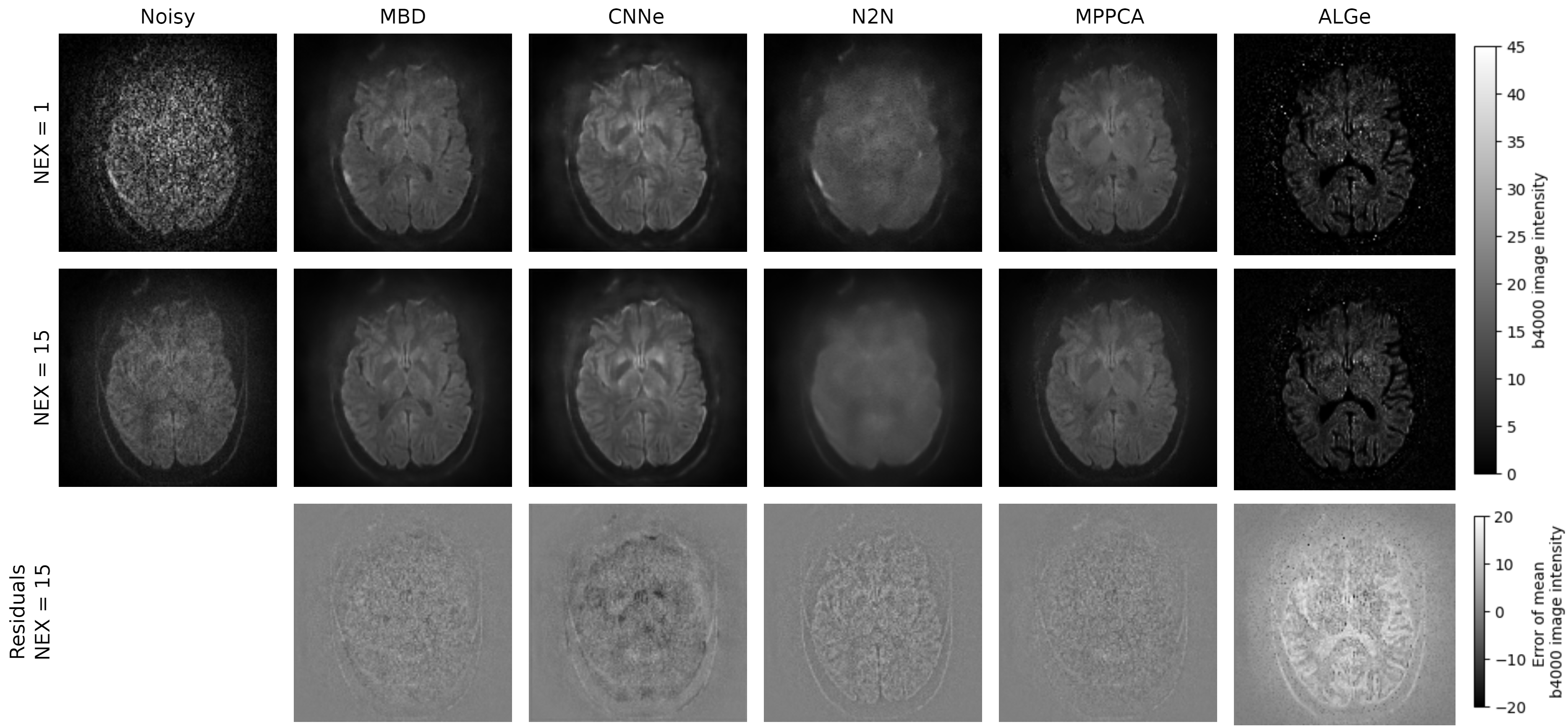}}\hfill
\caption{Comparison of MBD, CNNe, N2N, MPPCA and ALGe denoising of b4000 STE brain dMRI. MBD shows better precision compared to averaging (NEX=15) and outperforms by far N2N in detail preservation. MBD also avoids systematic errors visible in the CNNe results. MAE evaluation (Section~\ref{s:stedeno}) showed smaller error for MBD than for MPPCA. Missing brain details are visible in NEX=1 MPPCA-denoised image, for example in the occipital horn of the lateral ventricle on the left image side.}
\label{steimages}
\end{figure*}

Visual evaluation of the predicted b4000 images can be done based on Figure~\ref{steimages}. CNNe results are biased, with large areas of falsely elevated signal, as visible in the denoised images and the residual maps. N2N features blurring, which removed most anatomical details. MBD, in turn, is the most reliable among the learning-based methods. ALGe result is noisy and shows largest differences compared to reference. In comparison with MPPCA, MBD demonstrates similar visual quality when evaluated on averaged images, although this comparison is challenging due to the relatively high noise variance in the average of noisy images. On the other hand, the single b4000 image denoised using MPPCA shows blurring in certain slices. In the slice visualized in Figure~\ref{steimages}, this is visible in the unclear representation of the occipital horn of the lateral ventricles in the left half of the image, as well as the lack of some cortical sulci. These details are better represented in the averaged MPPCA image.

In Figure~\ref{steimages}, an area of elevated signal is visible in brain parenchyma, on the bottom left side of the image. This is not visible in the mean image, nor the CNNe and ALGe predictions. The averaged MBD image does not show it. We identified this artefact as imperfect fat saturation, so it is not induced by the method.

Quantitatively, MBD training resulted in the smallest validation loss (Supplement Figure~S6). Image intensity MAE computed on the test volume was 2.06, 0.74, 1.11, 0.84, 0.79 and 3.92 for noisy, MBD, CNNe, N2N, MPPCA and ALGe unaveraged image, in reference to the averaged noisy image. 


\subsection{Configurations of MBD}

%
%

\begin{figure*}[ht!]
\centering
\subfloat{\includegraphics[width=\textwidth]{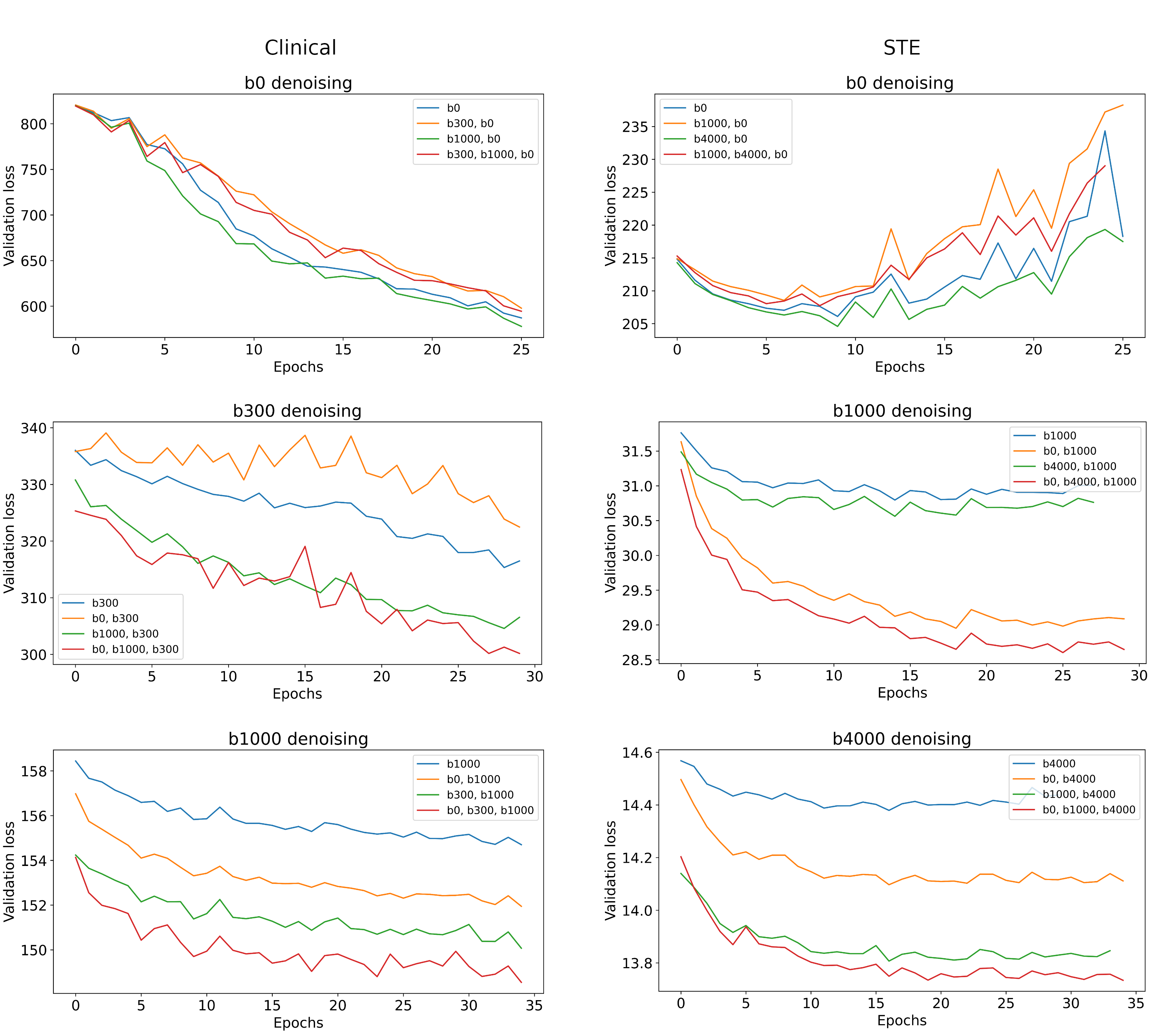}}\hfill
\caption{Validation (MSE) loss curves for different sets of input images for MBD. For denoising of images at b-values higher than 0, using all available b-value data was superior. For b0 images, a combination of the b0 and the highest available b-value was superior.}
\label{mbdcomparisonclinical}
\end{figure*}

Figure~\ref{mbdcomparisonclinical} shows denoising results for each individual b-value in the clinical and STE datasets. It reveals that combining images at multiple b-values at network input improves prediction regardless of the b-value of the predicted image. b0 prediction benefited from combining the b0 image with the highest b-value image at input, both for STE and clinical data. It was slightly, but consistently better than inputting just the b0 image (which is equivalent to N2N training). For two higher b-values of each of the triplets, denoising was superior when all b-values were used as input. 

We observed that the order of the input types, according to the minimal loss value, was the same for b0 and highest b-value image denoising in both brain dMRI datasets, but for the middle b-value, we observed differences between the STE and clinical data. For denoising of clinical b300 images, the worst input was b0 and b300 together, while the option of using the b0 and b1000 image was second best for STE data. The order of other combinations was the same, however, the b300 and b1000 input for clinical b300 denoising was almost as good as the best, while in STE it was just slightly better than the worst result. This effect could appear due to the closeness of b-values, but denoising of STE b1000 gave different results than of the b1000 image from the clinical dataset. We did not conclude on the reason, as more experiments are needed in this regard.

\subsection{Denoising synthetic dMRI with lesions}

The total MSE validation loss was in favour of MBD (79.88), lower than N2N (80.72) and CNNe (80.55), in reference to the theoretical minimum (79.63). For lesion voxels, however, CNNe loss was close to MBD. Lesion loss was higher than healthy tissue loss for all methods (Supplement Figure~S7). 

\begin{figure*}[ht!]
\centering
\subfloat{\includegraphics[width=\textwidth]{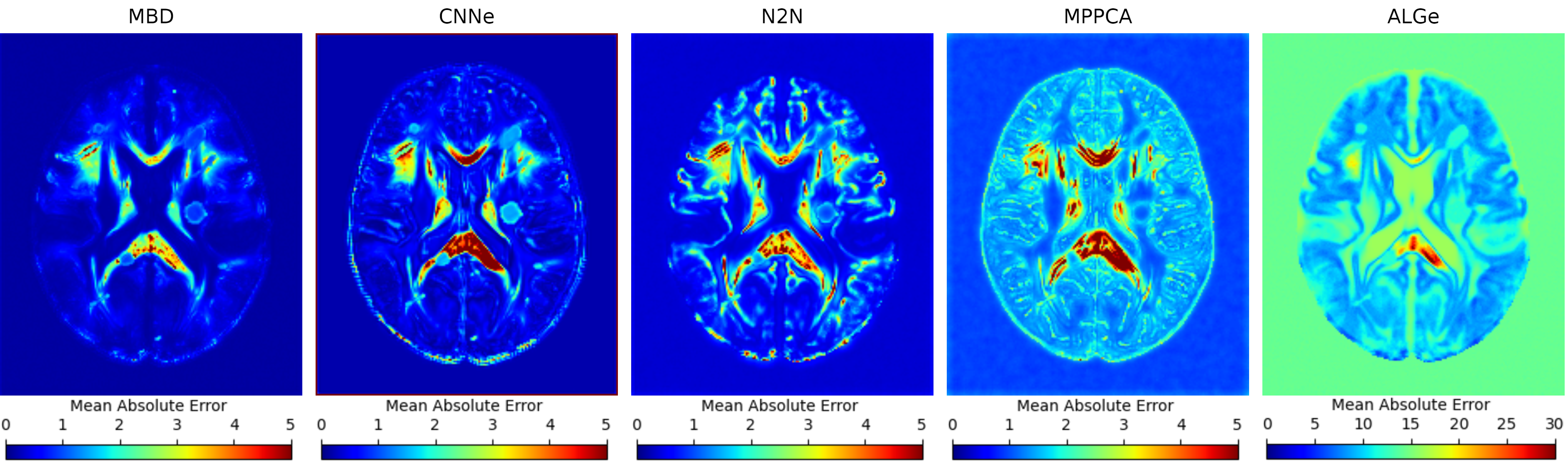}}\hfill
\caption{Average mean absolute error (expressed in DW image intensity units) maps for a selected slice. The average was computed across 1000 instances of the slice each with random lesion parameters. MBD shows lowest lesion mean absolute errors and lowest mean absolute errors in healthy tissue. MBD also showed good detail preservation accross the presented slice.}
\label{lesion1}
\end{figure*}

Figure~\ref{lesion1} shows mean absolute error maps computed over 1000 test slice repetitions with random lesion parameters. This graph reveals the systematic errors of each method. Denoised images show increased error in areas where intensity was high in b4000 images. ALGe was unable to extrapolate the b4000 signal reliably. MPPCA was better, but features more error than any of the learning-based methods. Among these, N2N showed high error on edges, related to blurring. MBD features lowest error values in the normal tissue and in the lesions. Error is elevated at lesion boundaries, which could result from partial volume effects. 

\begin{figure*}[ht!]
\centering
\subfloat[Intensity error, bins of size 0.01]{\includegraphics[width=0.5\textwidth]{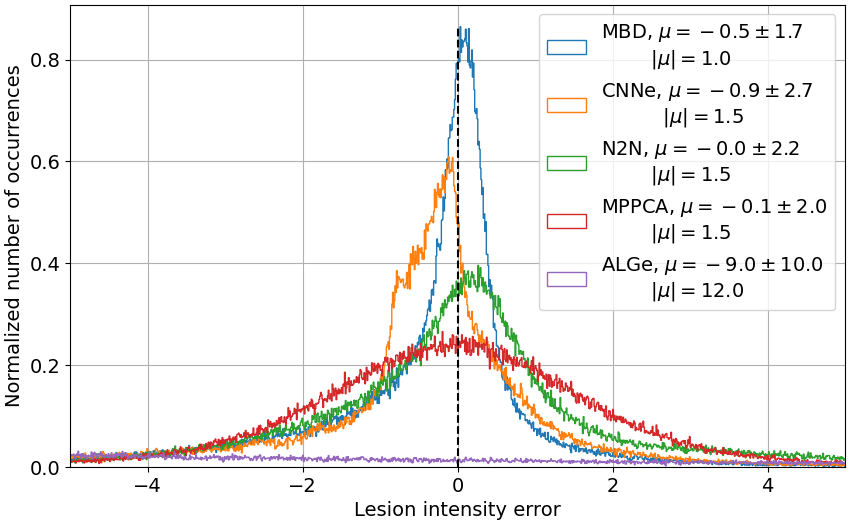}}\hfill
\subfloat[Absolute intensity error, bins of size 1]{\includegraphics[width=0.5\textwidth]{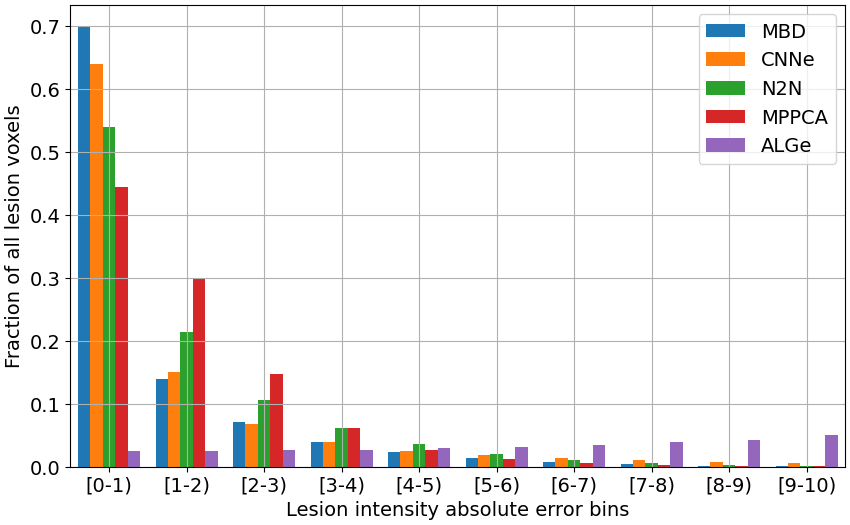}}\hfill
\caption{Normalized histograms for 1000 random lesion parameter sets and all test lesions. In (a), deviation of the mean from the 0 value shows that all methods tend to underestimate b4000 intensity slightly, with lowest shift for N2N and highest for CNNe, and smallest standard deviation for MBD. The mean absolute error $|\mu|$ is lowest for MBD. (b) shows that for all methods, errors are mostly in the interval [0,1), but MBD has the largest fraction of these, reaching 0.699.}
\label{lesion2}
\end{figure*}

The observations on lesions are supported by the lesion intensity error histograms. Histograms in Figure~\ref{lesion2} represent only those voxels which contain lesion tissue solely, across the 1000 repetitions with random lesion parameters. MBD has the tallest histogram, with the mode nearly at 0, and with the smallest standard deviation. It also has the smallest mean absolute error value of 1.5, compared to 1.5 achieved by CNNe, N2N and MPPCA. MPPCA has the most symmetrical error histogram, but it features a large standard deviation. MBD had over 69.9\% of voxel intensity absolute errors in the range of [0,1], while CNNe had 64.0\%, N2N 54.0\%, MPPCA 44.4\% and ALGe 0.02\% as shown in the first group of bars in Figure~\ref{lesion2}b. This shows that MBD errors are low more frequently than for the other methods.
 
\begin{figure*}[ht!]
\centering
\subfloat{\includegraphics[width=\textwidth]{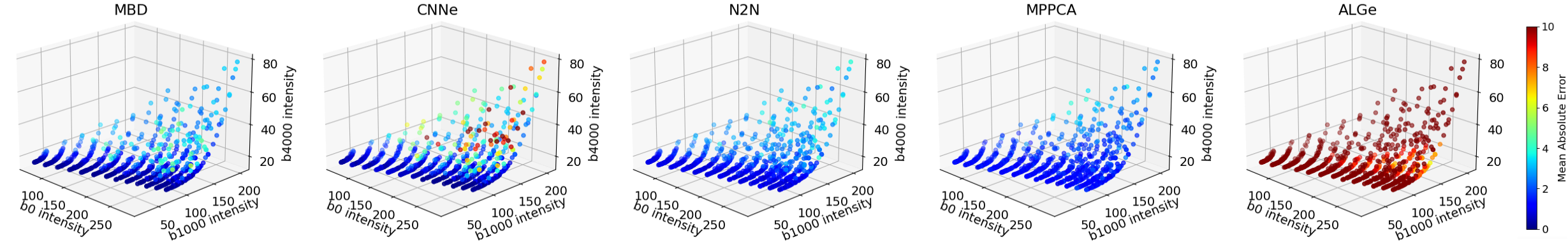}}\hfill
\caption{Scatter plot of lesion intensities with colour-coded mean absolute errors. CNNe errors reach higher values than MBD and N2N. For all methods, errors increase as image intensity increases. For comparison of individual error values across methods, see Figure~\ref{lesion4}.}
\label{lesion3}
\end{figure*}

Factors influencing errors, other than lesion shape and location (neighbourhood), are  voxel intensities across the images in the DWI sequence. Figure~\ref{lesion3} shows the distribution of b0, b1000 and b4000 image intensities, where each lesion parameter set (including different lesion shapes and positions) is represented by a dot. Colours denote mean absolute errors for the lesion voxels. This graph reveals that larger errors are associated with higher lesion intensity in b1000 and b4000 images. 

\begin{figure}[ht!]
\centering
\subfloat{\includegraphics[width=0.5\textwidth]{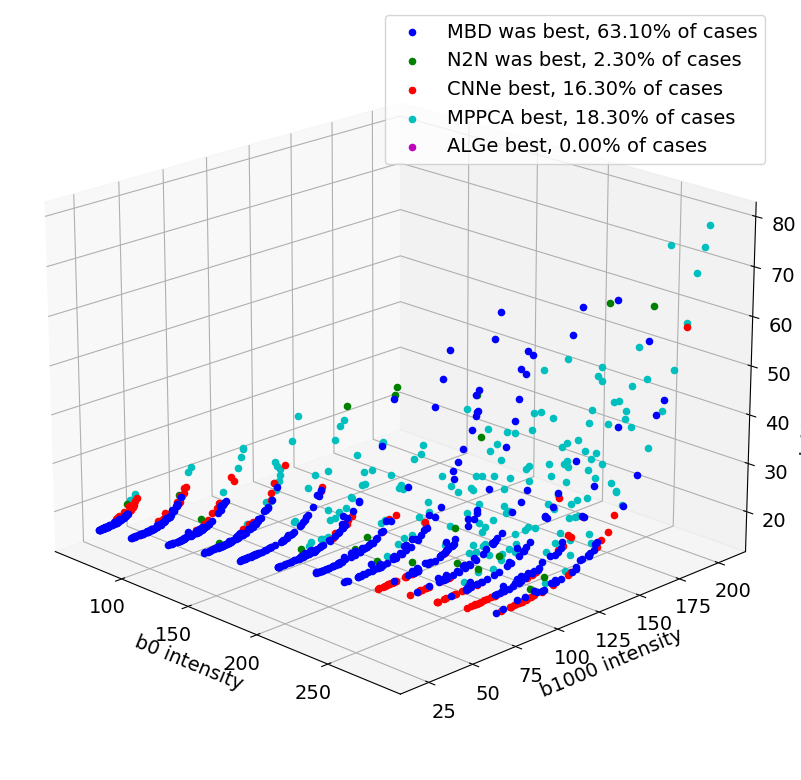}}\hfill
\caption{Scatter plot of lesion intensities, colours code the method which achieved the lowest mean absolute error. MBD is superior for more than 63\% of lesion parameter sets.}
\label{lesion4}
\end{figure}

For easier comparison between methods, Figure~\ref{lesion4} shows the same space with colours coding the method with lowest error. According to this criterion, MBD was best for over 63\% of lesion parameter sets, while the second best was MPPCA with 18.3\% of lesions cases denoised with smallest mean absolute error compared to other methods. Cases in which MPPCA was better than MBD align with the cases in which MBD had highest errors, i.e. for relatively high b4000 intensities.

\begin{figure*}[ht!]
\centering
\subfloat{\includegraphics[width=\textwidth]{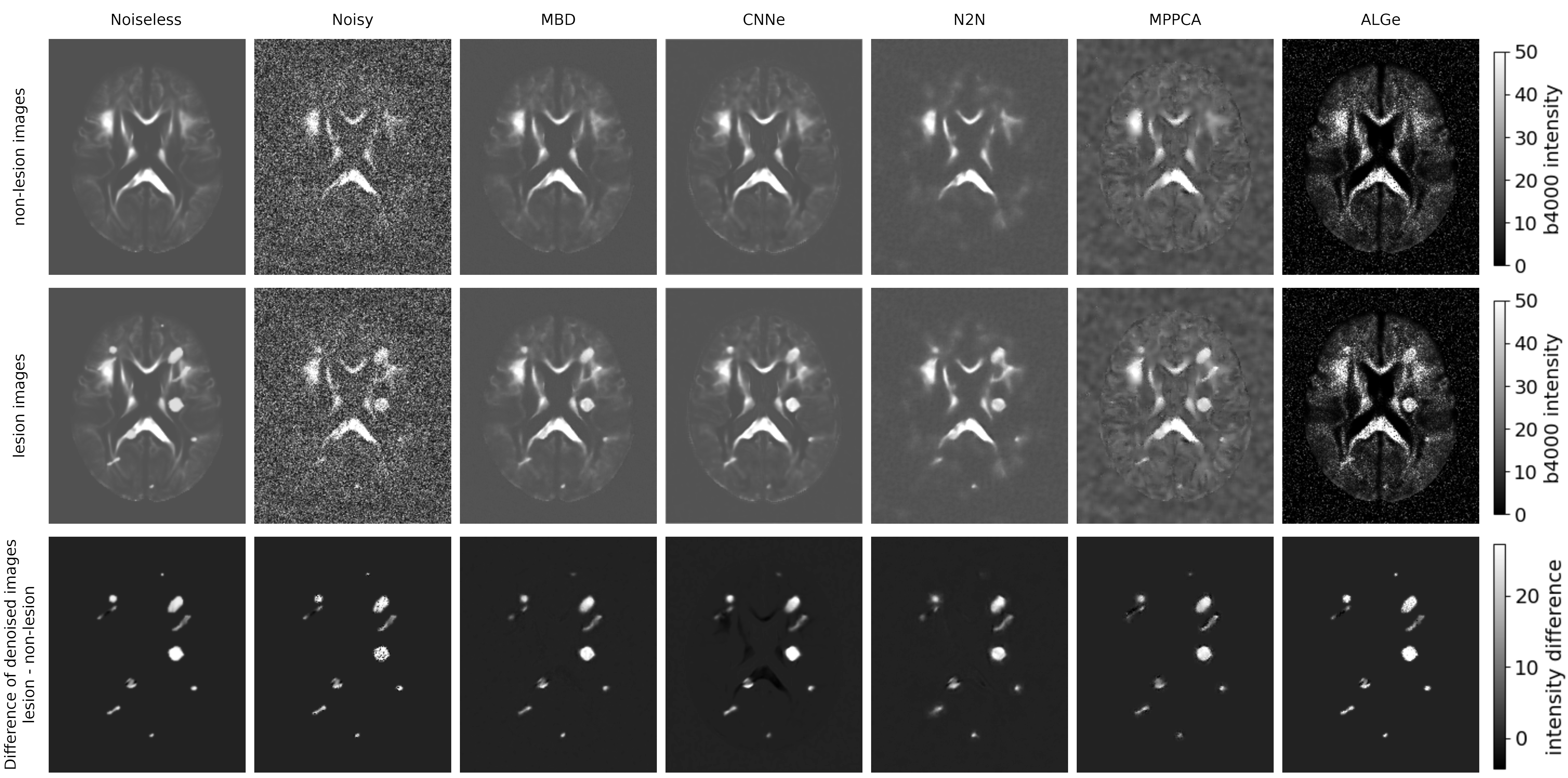}}\hfill
\caption{Comparison of synthetic dMRI denoised using MBD, N2N and CNNe. For each method, the lesionless test slice (upper row) and the corresponding slice with lesions (middle row) were denoised. Differences between both denoising results (bottom row) reveal lesion areas the consecutive networks indirectly detected. Lesion parameters:  $f=0.75$, $D_1=400\frac{mm^2}{s}$, $D_2=600 \frac{mm^2}{s}$, $\Delta T_2=15$ ms. See Figure~\ref{f:dtidata} for all b-values for that slice. All methods 'detect' lesions and preserve them in denoising. In general, MBD and CNNe show highest visual similarity to the noiseless reference, but the former estimates lesion intensity more accurately. More detailed comparison between MBD and other methods, in different lesion parameter sets, is shown in Figures~\ref{f:lesionset1}-\ref{f:lesionset3} and in Supplement Figures~S8-S11.}
\label{lesion0}
\end{figure*}

Difference maps between lesion and lesionless images for a chosen lesion parameter set, which yield hyperintensive lesions across all investigated b-values, are shown in Figure~\ref{lesion0}. For any chosen method, the general appearance of the denoised images is the same independent of the presence or absence of lesions. Consequently, lesions conspicuity remains unchanged for each of the methods, which is visible in the lower row of Figure~\ref{lesion0}. For N2N, MPPCA and ALGe, the denoised images are visibly different from the ground truth. MBD and CNNe images are very similar visually, despite the quantitative difference discussed previously. However, CNNe seems to have unexpected errors outside lesions, in white matter regions of higher intensity, caused solely by the occurrence of the lesions in the image. Such effects are also visible for MBD and N2N, but to a negligible extent.

\begin{figure}[ht!]
\centering
\subfloat{\includegraphics[width=0.19\columnwidth]{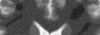}}\hfill
\subfloat{\includegraphics[width=0.19\columnwidth]{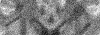}}\hfill
\subfloat{\includegraphics[width=0.19\columnwidth]{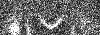}}\hfill



\subfloat{\includegraphics[width=0.19\columnwidth]{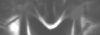}}\hfill

\subfloat{\includegraphics[width=0.19\columnwidth]{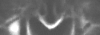}}\hfill
\subfloat{\includegraphics[width=0.19\columnwidth]{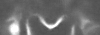}}\hfill
\subfloat{\includegraphics[width=0.19\columnwidth]{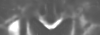}}\hfill
\subfloat{\includegraphics[width=0.19\columnwidth]{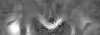}}\hfill
\subfloat{\includegraphics[width=0.19\columnwidth]{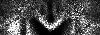}}\hfill

\subfloat{\includegraphics[width=0.19\columnwidth]{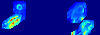}}\hfill
\subfloat{\includegraphics[width=0.19\columnwidth]{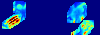}}\hfill
\subfloat{\includegraphics[width=0.19\columnwidth]{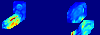}}\hfill
\subfloat{\includegraphics[width=0.19\columnwidth]{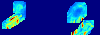}}\hfill
\subfloat{\includegraphics[width=0.19\columnwidth]{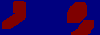}}\hfill

\subfloat{\includegraphics[width=0.19\columnwidth]{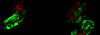}}\hspace{0.01\columnwidth}
\subfloat{\includegraphics[width=0.19\columnwidth]{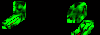}}\hspace{0.01\columnwidth}
\subfloat{\includegraphics[width=0.19\columnwidth]{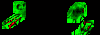}}\hspace{0.01\columnwidth}
\subfloat{\includegraphics[width=0.19\columnwidth]{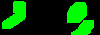}}

\caption{Denoising of the synthetic dMRI with lesions, $f=0.65$, $D_1=400\frac{mm^2}{s}$, $D_2=700 \frac{mm^2}{s}$, $\Delta T_2=-20$ ms. The top row shows the b0, b1000, b4000 sequence of noisy DWI. Second top row shows the clean reference image. The three lower rows show denoising results (from left: MBD, N2N, CNNe, MPPCA, ALGe), error maps masked for lesions (from left: MBD, N2N, CNNe, MPPCA, ALGe), maps of error difference (from left: MBD vs. CNNe, MBD vs. N2N, MBD vs. MPPCA, MBD vs. ALGe). Shades of green in the error difference maps mean MBD had lower error, shades of red mean MBD had higher error. MBD shows better denoising for most lesion voxels as compared to other methods.}
\label{f:lesionset1}
\end{figure}

\begin{figure}[ht!]
\centering
\subfloat{\includegraphics[width=0.19\columnwidth]{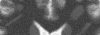}}\hfill
\subfloat{\includegraphics[width=0.19\columnwidth]{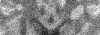}}\hfill
\subfloat{\includegraphics[width=0.19\columnwidth]{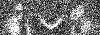}}\hfill
%
%

\subfloat{\includegraphics[width=0.19\columnwidth]{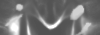}}\hfill

\subfloat{\includegraphics[width=0.19\columnwidth]{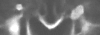}}\hfill
\subfloat{\includegraphics[width=0.19\columnwidth]{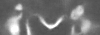}}\hfill
\subfloat{\includegraphics[width=0.19\columnwidth]{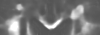}}\hfill
\subfloat{\includegraphics[width=0.19\columnwidth]{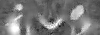}}\hfill
\subfloat{\includegraphics[width=0.19\columnwidth]{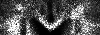}}\hfill

\subfloat{\includegraphics[width=0.19\columnwidth]{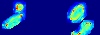}}\hfill
\subfloat{\includegraphics[width=0.19\columnwidth]{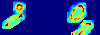}}\hfill
\subfloat{\includegraphics[width=0.19\columnwidth]{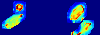}}\hfill
\subfloat{\includegraphics[width=0.19\columnwidth]{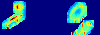}}\hfill
\subfloat{\includegraphics[width=0.19\columnwidth]{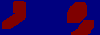}}\hfill

\subfloat{\includegraphics[width=0.19\columnwidth]{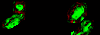}}\hspace{0.01\columnwidth}
\subfloat{\includegraphics[width=0.19\columnwidth]{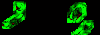}}\hspace{0.01\columnwidth}
\subfloat{\includegraphics[width=0.19\columnwidth]{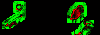}}\hspace{0.01\columnwidth}
\subfloat{\includegraphics[width=0.19\columnwidth]{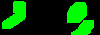}}

\caption{Denoising of the synthetic dMRI with lesions, $f=0.65$, $D_1=400\frac{mm^2}{s}$, $D_2=700 \frac{mm^2}{s}$, $\Delta T_2=20$ ms. The top row shows the b0, b1000, b4000 sequence of noisy DWI. Second top row shows the clean reference image. The three lower rows show denoising results (from left: MBD, N2N, CNNe, MPPCA, ALGe), error maps masked for lesions (from left: MBD, N2N, CNNe, MPPCA, ALGe), maps of error difference (from left: MBD vs. CNNe, MBD vs. N2N, MBD vs. MPPCA, MBD vs. ALGe). Shades of green in the error difference maps mean MBD had lower error, shades of red mean MBD had higher error. MBD shows better denoising for most lesion voxels as compared to other methods.}
\label{f:lesionset3}
\end{figure}

Lesion denoising quality is further studied for chosen parameter sets, including lesions that were hypo-, iso- and hyperintense in b0 and lesions that appeared hypo-, iso- or hyperintense in b4000. Figures~\ref{f:lesionset1}-\ref{f:lesionset3} show the test slice, cropped to the nearly frontal brain area, which shows four lesions: a medium-sized round, two elongated with medium width and a big oval. Error maps were masked to focus on regions with a non-zero content of lesion tissue. To avoid random errors resulting from specific noise instances, the computed error maps are an average of 100 repetitions.

ALGe has highest errors, higher in all voxels than MBD. N2N and MPPCA show higher error than MBD and CNNe for the large isointensive lesion in Figure~\ref{f:lesionset1}. Foci of elevated error in lesions are very similar for MBD and CNNe. To enable a clearer comparison of error maps, we computed error difference maps for MBD-CNNe, MBD-N2N, MBD-MPPCA and MBD-ALGe pairs. These maps are shown in pseudocolour in the lower rows of Figures~\ref{f:lesionset1}-\ref{f:lesionset3}, with the green colour denoting superiority of MBD, and red denoting superiority of the other method. MBD achieved lower error than other methods for most lesion voxels regardless of the parameter set. The worst result for MBD was obtained for the lesions that were hypointense in b0, b1000 and nearly isointense in b4000 (Figure~\ref{f:lesionset1}). Results for other parameter sets are presented in the Supplement, Figures~S8-S11.

\section{Discussion}

\subsection{Main findings and relation to other methods}

We proposed a data-efficient learning approach to diffusion MRI denoising, called MBD. It shares the concept of feeding additional images to a neural network to facilitate denoising. In dMRI, one of the possible forms of guided denoising is feeding diffusion-weighted images at different b-values. Thus, we proposed a more specific name to this variation of guided denoising, multi-b-value-based denoising or MBD. Up to our knowledge, this idea was first applied to dMRI by Kaye et al.~\cite{Kaye2020} for the prostate, then reused in rectal cancer~\cite{Mohammadi2023}, and recently in the prostate again~\cite{Pfaff2024}. 

Our work is significantly different from previous research on guided denoising in dMRI. We were initially unaware of the results in~\cite{Kaye2020, Mohammadi2023} and \cite{Pfaff2024} was published concurrently. We tested MBD on the previously unexplored brain imagery. We also incoporated phantom experiments on lesions with a biexponential signal model. In contrast to~\cite{Kaye2020, Mohammadi2023}, we use a single repetition image (NEX=1) as target, while they used averaged images. Using averaging promotes motion-related blurring and thus it may lead to a low pass behaviour of the neural network. Unlike~\cite{Pfaff2024}, who focused on self-supervised denoising, we use a separate training target. We therefore do not rely on the accuracy of phase correction and noise variance estimation that could influence the quality of denoising and which was inevitable in~\cite{Pfaff2024}. We also expanded the scope of application of this methodology and investigated the risks and benefits from guidance in dMRI in more detail.

We showed MBD can deal with healthy brain images with b-value up to 4000 s/mm$^2$, significantly improving signal precision and preserving edges. Variance reduction, bias reduction and edge-preservation are among the most important features of an ideal denoising method~\cite{ManzanoPatron2024}. We used a fixed neural network architecture to demonstrate that reinforcing the neural network input improves edge-preservation, especially for denoising low SNR images. For b4000 brain images, MBD revealed more details than an average over 15 repetitions, while N2N featured excessive blurring. This comparison also held for b1000 images, although the difference was not as striking. We therefore assume that N2N works similarly well to MBD for high SNR images, but the latter is much better at low SNR.

The edge-preserving feature of MBD could be explained by the SNR characteristics of a DWI dataset. This explanation was suggested in~\cite{Kaye2020}. At lower b-values, SNR is higher and thus the structure of the object is represented with higher quality. This representation is observed by the neural network, extracted and implemented into the denoised high b-value image. However, this quality of the method cannot be attributed solely to the mentioned characteristics. It was demonstrated that denoising DWI at intermediate b-values in the investigated datasets also benefited from using all available b-values, including higher ones. With this in mind, we assume the the neural network exploits the implicit signal model. We treat this as an important property of the method, since this signal model comes from data, so it is not simplified as the explicit monoexponential model, for example. 

Unexpectedly, denoising of b0 images benefited from including just the highest available b-value at the network input. The latter effect remains unexplained, but the experiment shows that infomation necessary for denoising of a diffusion MR image at some b-value can also be found at higher b-values. In this light, we suggest that MBD observes both the spatial and b-value-space context and thus it makes better predictions.

We are aware that neural networks pose a potential risk of displaying visually pleasing images that are actually do not represent the object accurately. In the specific context of the denoising problem, one risk might be that when denoising very low SNR images, like b4000, the network may excessively use lower b-value images and extrapolate. We therefore thoroughly studied a network which did not observe b4000 images (CNNe) and one that observed all available b-values (MBD). Indeed, CNNe showed pleasing visual results, manifesting themselves in the good resolution of the predicted images, both in-vivo and in the phantom. However, quantitative evaluation using validation loss showed that MBD was closer to the reference than CNNe, proving that MBD is not using undesired extrapolation and information from the high-b-value image was used to improve the accuracy. In other words, pure extrapolation cannot account for the high performance of MBD. In visual examination, CNNe extrapolation of b4000 images yielded clearly biased images, in contrast to MBD. The bias was systematic as it was not reduced by averaging. The bias was even more obvious for the clinical dataset at b1000. We thus stress that while CNNe predictions are visually convincing, this method is not recommended for quantitative data interpretation due to a large bias. A question that remains without answer is if CNNe would have reduced bias if more b-values were used. We speculate that increasing the b-value range would be beneficial in that regard, but not just increasing the number of acquired b-values. Taking into account that algebraic extrapolation yielded the worst results in all experiments and requires averaging to improve precision, CNNe might be used to replace simple extrapolation in scenarios when MBD is not possible. It could be used by radiologists in prostate diffusion imaging, where extrapolation of high-b-value images is used. Although this must be investigated more systematically, our experiments suggest that extrapolation-based images may suffer large errors. Therefore, both ALGe and CNNe approaches seem unreliable for quantitative interpretation.

We used syntethic DWI with lesions to verify if MBD is able to deal with structures that may have a different underlying diffusion model that healthy tissue. In our experiments, we assumed healthy white matter was monoexponential and lesions were biexponential. For a large set of different lesion diffusion parameters, we found that MBD had lowest error in 63\% of cases. This shows that the approach of MBD could become a clinically useful tool. However, more experiments using patient data are required to verify this.

In addition to algebraic extrapolation and other supervised learning methods, we included MPPCA as another reference method. Our experiments were not prepared to provide an extensive comparison between MBD and MPPCA. MPPCA, originally proposed for multidirectional dMRI, is one of the state of the art dMRI denoising methods with excellent performance on data with high levels of redundancy. Thus, deploying it on data with low redundancy is intended only as a benchmark comparison that highlights the potential benefit of methods adapted to small sets of data, such as MBD. In the STE case, MPPCA showed quantitative results close to MBD. Visually, MPPCA resulted in local contrast of detail loss. This effect was probably random, since those losses are not visible in the average of denoised images. However, MBD readily avoids them before averaging and as such it does not require multiple scan repetitions. We observed some intensity differences between the averaged MBD- and MPPCA-denoised images. Specifically, a hyperintense region is visible below the globus pallidus on the left side of the image, in Figure~\ref{steimages}, for the NEX=15 MBD image. This spot can be recognized in the averaged residuals and thus can be treated as a denoising mistake that MPPCA avoided. This spot is hyperintense in the b1000 image, too. It is likely an anatomical structure, since it spans accross multiple slices when the whole image volume is viewed, but may have incorrect b4000 intensity after denoising. This 'shine-through' effect could possibly be avoided if the b4000 image was originally at higher SNR. That could be achieved by prior averaging, for example with NEX=2. The underlying tissue might also be an outlier in the training set, for which not enough training examples was available. In that case, increasing the number of patients in the training set would be helpful. It is also worth to emphasize that MPPCA, for the in-vivo experiments, was allowed to use 3D context, while MBD was limited to the denoised 2D slice. Despite the possible mistake, the average of MBD-denoised images was closer to the average of noisy ones than the average of MPPCA-denoised images. Apparently, there was more intensity inconsistency in MPPCA-denoised images in the whole image volume.
In the clinical case, MPPCA was allowed to use three diffusion directions (orthogonal, thus with minimal redudancy) and three b-values, while MBD used only one diffusion direction as input. In this case, MBD had higher precision improvement. From this limited comparison, we conclude that MBD is a competetive method. We also notice that MBD is generally more computationally efficient. 

\subsection{Study limitations}

The current study features a few limitations that will be addressed in the future research. In our view, the main limitation was the lack of an in-vivo dataset with tumour data. For this reason, we simulated lesions using a brain phantom. Simulated lesions where homogeneous to some extent, which could make them easier to denoise using methods like N2N, which featured blurring. On the other hand, we were able to simulate a large number of tissue parameters for the same lesion shapes and locations, which allowed to compute the average error. It was shown on artificial lesions for MBD, that higher errors were made for brighter lesions in b4000 images. This conclusion is surprising as brighter lesions have higher SNR and thus should be easier to denoise. We suspect this issue may be related to the training set imbalance, where low intensities dominate. Alternately, it could be attributed to the oversimplified brain phantom morphology model used together with a more realistic diffusion tensor distribution, which has prevalent impact on the contrast at high b-values. The effect was not only observed for lesions, but also for the healthy, monoexponential tissue. Despite the phantom experiment showed that MBD and other methods can deal with lesions, which were modelled as a bicompartmental tissue in contrast to the healthy white matter, in-vivo experiments are necessary to confirm our the observations. 

An interesting, but unexplored scope of research is the performance of MBD in datasets with high directional data redundancy. In such a setting, MPPCA is one of the state of the art methods. We will study if MBD could be competetive to redundancy-based methods in such a scenario. 

In the current study, we focused on brain images. However, MBD is in principle applicable to any body region. Indeed, MBD could be used for denoising of high-b-value prostate dMRI, where SNR is low and probing three diffusion-encoding directions is sufficient. This may unlock rapid and high-resolution dMRI in the prostate instead of current methods based on extrapolation, with poor accuracy, or averaging, with low speed. 

A drawback of MBD is that in the current form, it requires a sequence of diffusion-weighted images with matched diffusion-encoding direction, except for the b0 image. b-value shells in real world data usually use different diffusion directions between shells. This means that e.g. b300 data is acquired with different diffusion sampling directions than b1000 data, and, the numbers of directions per shell need not to be equal. The current version of MBD can deal with these data by inputting just two images at different b-values (i.e. b0 and bX). This can also be used for DTI data, where typically just b0 and b1000 images are acquired. However, as this problem could happen in a multidirectional dataset only, it could be solved by interpolating the image at a missing direction at an intermediate b-value from a small subset of other directions. This option could be applicable for DKI. Alternatively, all images in the intermediate b-value shell could be used as input instead of the desired direction and the network could learn to extract necessary information from them. These questions will be investigated in the future. 

We did not attempt to optimize the b-values used as input for denoising. We speculate that this choice has impact on the result. On the other hand, we showed that with three available b-values, denoising the image at the highest one always benefited from using all available data.

We focused on showing the benefits of denoising the images with highest b-values, since they have the lowest SNR. MBD, in principle, is not limited to such an application. We showed that b0 denoising and denoising of intermediate b-value images also benefited from the MBD approach. 

Finally, we focused on the idea of inputting multiple b-values to a neural network and not on the network architecture itself. We assume MBD results would improve for a much larger dataset and a larger network architecture. MBD could also be transformed into a multi-input and multi-output network, with simultaneous denoising of images at many b-values.

\section{Conclusion}

Multi-b-value-based denoising improves convolutional neural network-based dMRI noise removal results. The proposed method is edge-preserving and improved signal precision on a clinical brain dMRI and an ultra high b-value STE dMRI brain dataset with b=4000 s/mm$^2$. It is a promising tool for denoising, especially for datasets with a minimal number of diffusion encoding directions, such as clinical brain or prostate dMRI. 
\section{Code}
Python code that demonstrates a simple MBD application will be shared on GitHub in the near future.

\section{Acknowledgments}
The Authors thank Frederik Testud from Lund University/Siemens Healthcare AB for reviewing the manuscript and his useful comments.
We thank Siemens Healthineers (Forchheim, Germany) for providing sequence source and access to the pulse sequence programming environment.

\end{document}


\begin{frontmatter}
\title{Supplement to:\\MBD: Multi b-value Denoising of Diffusion Magnetic Resonance Images}%

\author{Jakub Jurek$^1$}
\author{Andrzej Materka$^1$}
\author{Kamil Ludwisiak$^2$}
\author{Agata Majos$^3$}
\author{Filip Szczepankiewicz$^4$}

\address{$^1$Institute of Electronics, Lodz University of Technology, Aleja Politechniki 10, PL-93590 Lodz, Poland}
\address{$^2$Department of Diagnostic Imaging, Independent Public Health Care, Central Clinical Hospital, Medical University of Lodz, Pomorska 251, PL-92213 Lodz, Poland}
\address{$^3$Department of Radiology, Medical University of Lodz, Lodz, Poland}
\address{$^4$Medical Radiation Physics, Lund University, Barngatan 4, 22185 Lund, Sweden}

\end{frontmatter}

\section{Supplementary methods}
\subsection{Extraction of lesions for simulation from a real dataset}

\begin{figure}[ht!]
\centering
\subfloat{\includegraphics[width=\columnwidth]{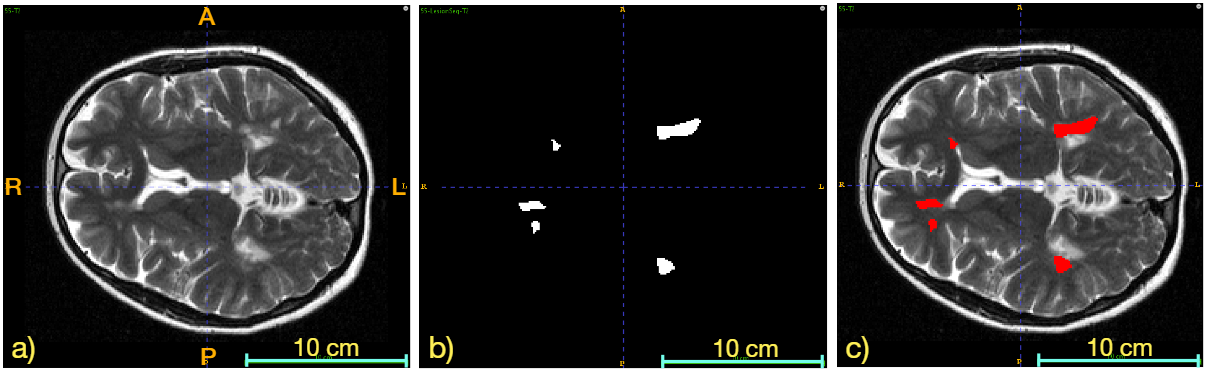}}\hfill
\caption{a) Example slice No 13 of patient No 55 T2w volume in dataset [1], b) binary masks of 5 lesions marked by three readers, c) the lesion masks superimposed on the image in a), in red.}
\label{f:lesion1}
\end{figure}
For sixty patients, the MS disease was diagnosed based on clinical criteria. The MS lesions were segmented in the coronal MRI slices by two radiologists and one neurologist. The consensus lesion regions were then made available as binary NIfTI volumes, along with the multi-sequence scans. From those, we selected the T2w dataset, from which 32 cases were chosen, featuring the slice thickness of 5 mm, the in-plane pixel resolution of 256x256, and the average spacing between slices of 5.5 mm. The number of slices a volume contains varies from 18 to 31 over this subset, with a median of 22. Figure~\ref{f:lesion1} shows an example of a coronal T2w slice No 13 for Patient No 55, along with the consensus MS lesions binary masks marked on this slice.

The average image intensity for a given brain tissue in the selected T2w images varies between patients, following various combinations of the repetition time and echo time applied for the measurement sequences. To make the extracted lesion images independent of those variations, we divided the pixel intensities inside lesion to the average intensity of the neighbouring white matter pixels. The procedure was as follows. An in-house data processing algorithm was implemented with the use of Python v. 3.10 numpy, scipy.ndimage, nibabel, pickle and matplotlib libraries. First, image slices containing lesions were identified, typically about 10 slices per patient. In each slice, the binary lesion masks were labelled. Some of the binary regions available in the segmented NIfTI slices of the dataset~\cite{Muslim2022} contain "holes" -- pixels of the background values. The hole filling procedure was then applied for each non-background region, followed by finding a rim composed of white-matter pixels closely surrounding the lesion, sort of an expanded lesion contour. The mean value of this contour intensity was used to normalize the extracted lesion regions. In effect, the lesion masks intensity was centred around the value of one. 
\begin{figure*}[ht!]
\centering
\subfloat{\includegraphics[width=\textwidth]{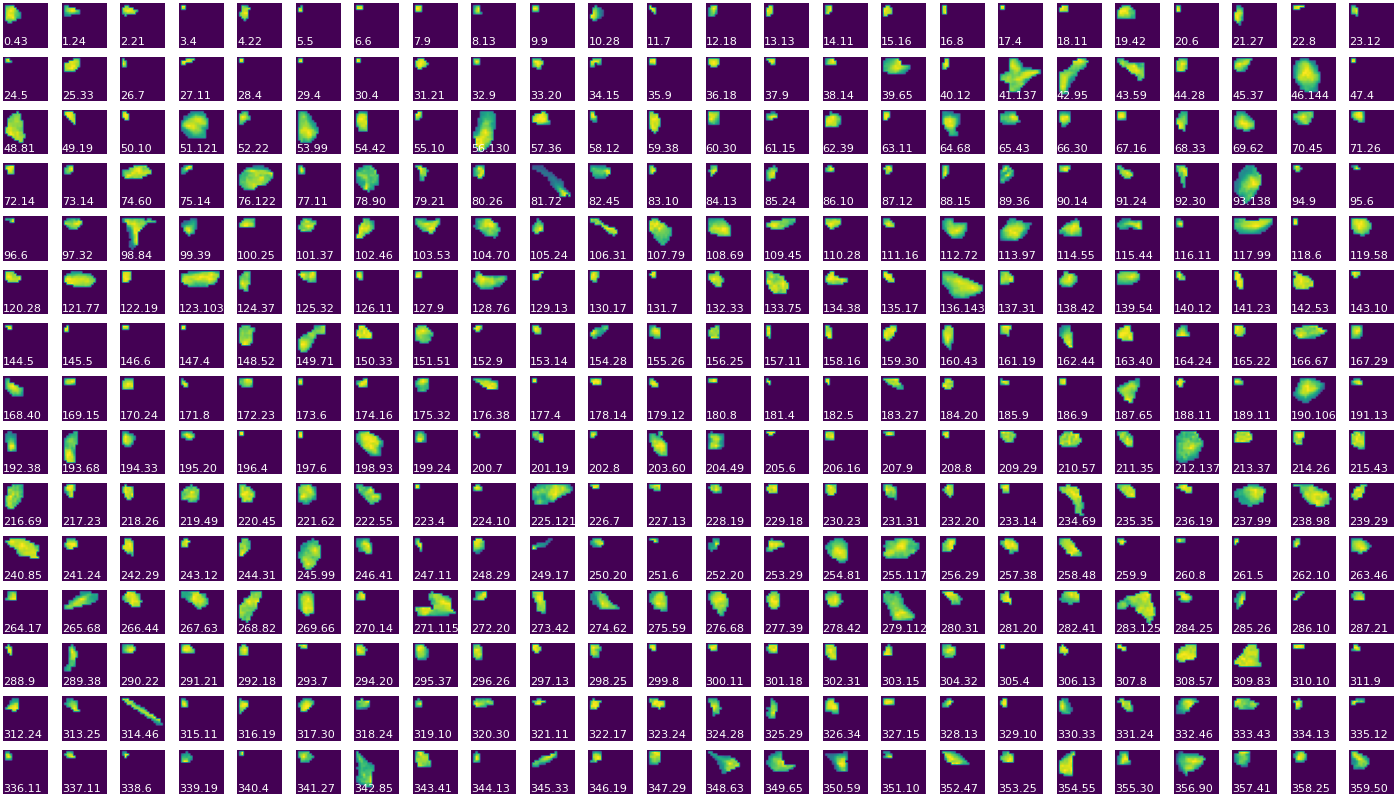}}

\subfloat{\includegraphics[width=\textwidth]{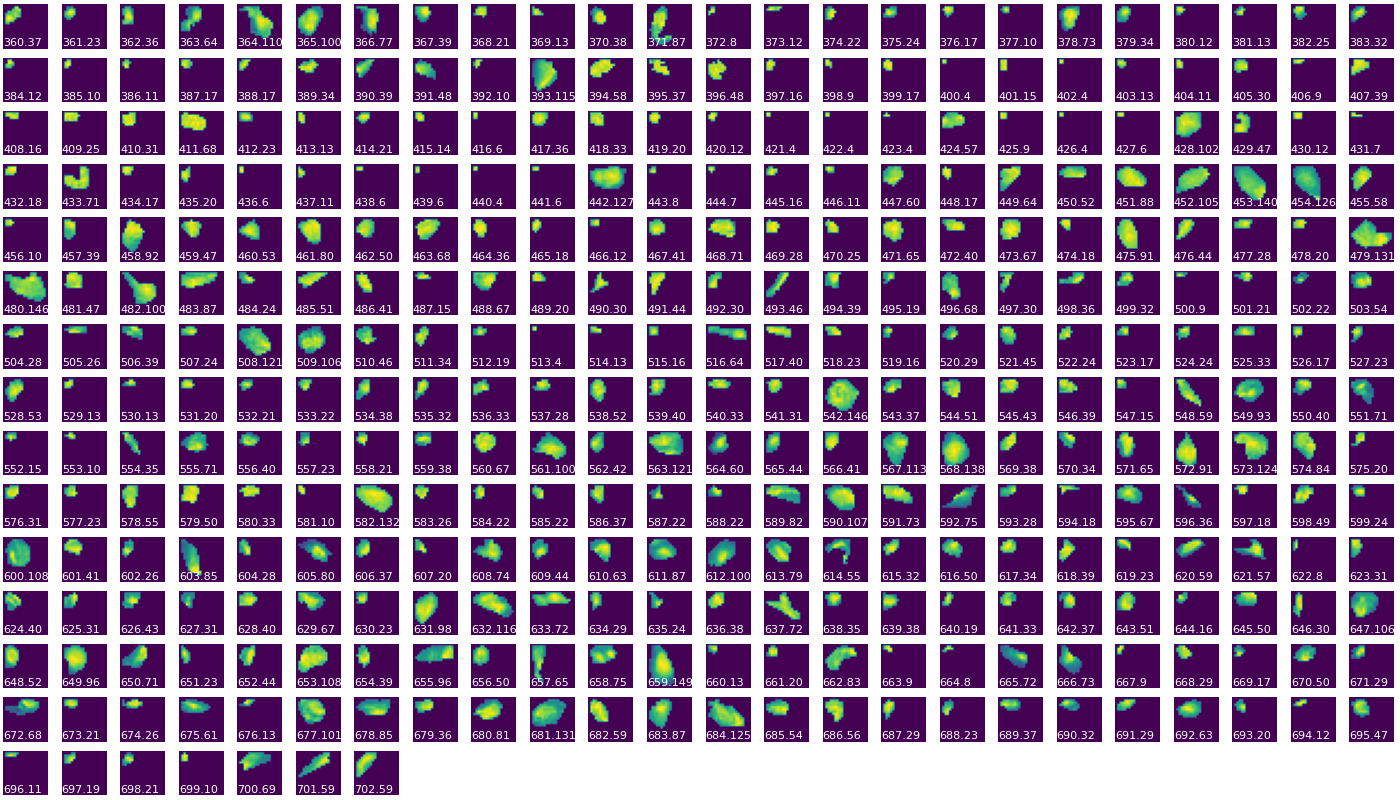}}
\caption{Mosaics of MS lesion masks of normalized intensity, extracted from the T2w MRI images of the dataset [1]. The numbers in the lower left corner of each tile, e.g. 702.59 denote lesion id number and its area in pixels. Upper panel: lesions from 0 to 359, lower panel: lesions from 360 to 702}
\label{f:lesion2}
\end{figure*}
\begin{figure}[ht!]
\centering
\subfloat{\includegraphics[width=\columnwidth]{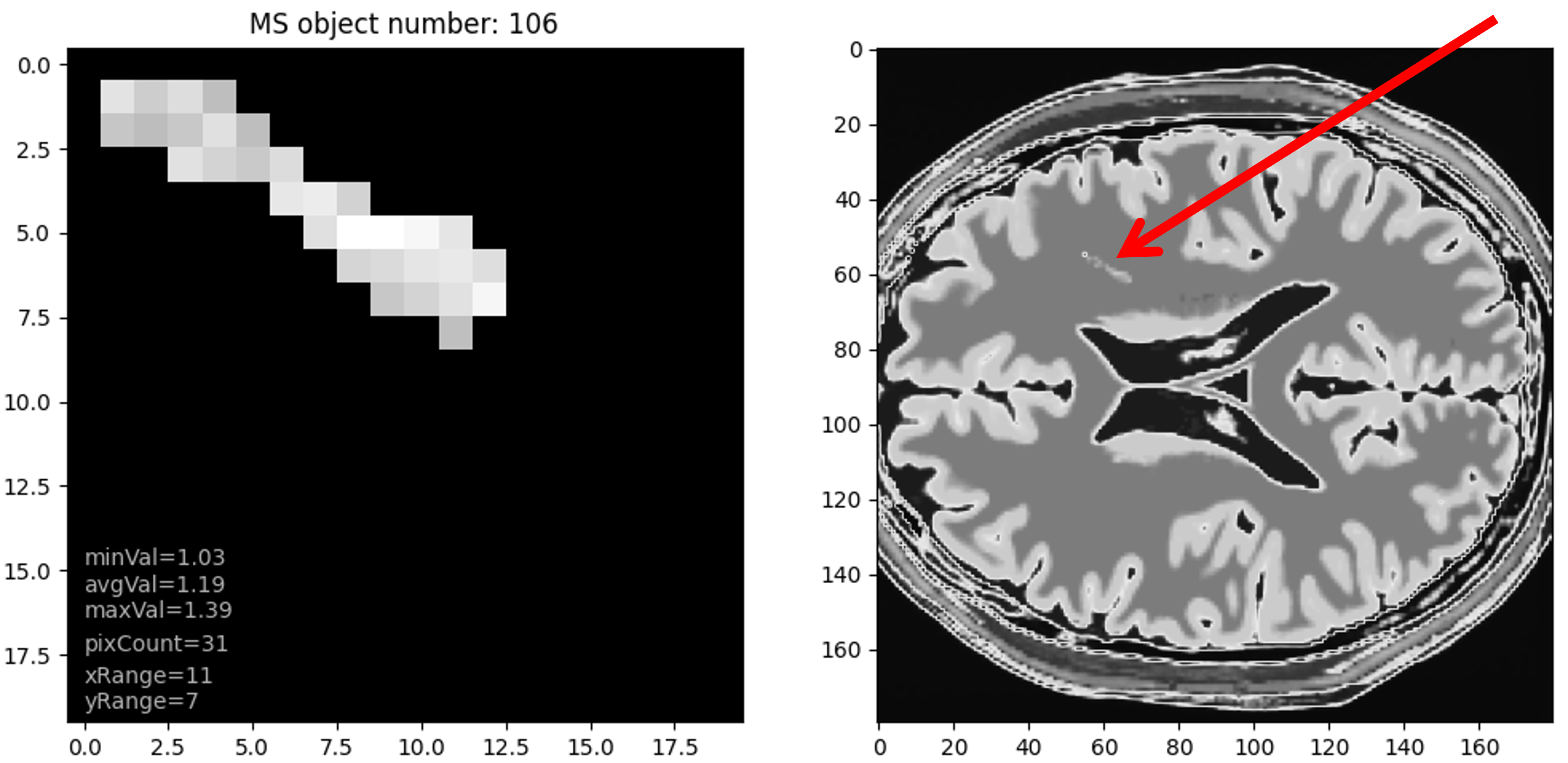}}\hfill
\caption{Left: An example of lesion mask No 106 containing 31 pixels of minimum, average and maximum intensity equal to 1.03, 1.19 and 1.39, respectively. This lesion (together with its expanded contour, not shown) is confined in the ((0,0),(13,9)) rectangle. Right: the lesion inserted into a synthesized T2w MRI, with its (0,0) pixels moved to (55,55) and intensity multiplied by the local WM value of 180}
\label{f:lesion3}
\end{figure}

For each lesion, a rectangle containing all expanded contour pixels was found. The lesions for which any side of their rectangle was shorter than 21 pixels and longer than 3 pixels were excluded from further consideration, leaving 703 regions extracted from the subset of 32 T2w volumes as presented in Figure~\ref{f:lesion2}. A data structure was designed to store the information needed to insert the lesions into the DWI synthetic images. The components of this structure are lists of lesion pixels intensity and their coordinates relative the encompassing rectangle (0,0) vertex, expanded contour coordinates, height and width of the rectangle and the average lesion intensity. The list of the structure values for all 703 lesion were stored as a Python .picle file. Figure~\ref{f:lesion3} shows and example of lesion No 106 inserted into the WM region of a synthesized T2w image slice. 
\begin{figure}[ht!]
\centering
\subfloat{\includegraphics[width=\columnwidth]{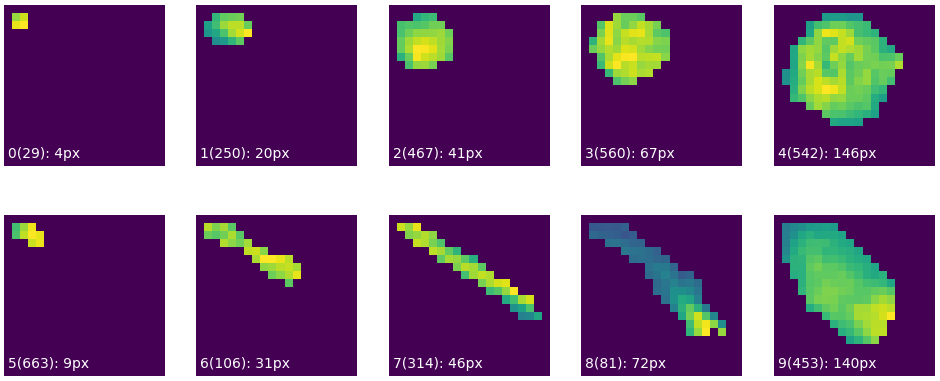}}\hfill
\caption{Test subset of MS lesion masks, used for evaluation of denoising algorithms}
\label{f:lesion4}
\end{figure}
Figure~\ref{f:lesion4} presents 10 lesion masks of different size, 5 of round shape and 5 elongated, chosen to form a test subset. The collection of the remaining 603 MS lesion masks formed the data for simulation of DWI brain slices in the training set. After lesion image preprocessing aimed at imprinting the diffusivity and other parameters distribution (explained below), they were randomly inserted into the WM masks of the IIT phantom MRI slices. For a given slice, the number of lesions was drawn from the range [4,...,10], depended on the WM mask area. The lesions were picked up randomly without replacement from the corresponding set (training or test). For any lesion, its location within the white matter region was randomly chosen, such that all lesion pixels lie within this region and the lesion mask does not overlap with any already inserted ones. 

Lesion masks were processed to obtain fuzzy lesion tissue maps consistent with the IIT phantom. Gaussian blurring with the sigma parameter of 0.7 was used to assure smooth transition of the lesion to the background. Lesion fraction values were then normalised to the [0,1] interval, by dividing by the factor of 3.5 mean lesion intensity and then by clipping intensities larger than 1. This flattens to a some extent the intensity of lesion depicted in Figure~\ref{f:lesion4}, but this way of normalisation allowed to obtain sufficiently many voxels with 100\% lesion fraction that could be later evaluated more easily. The sigma parameter for blurring and the normalisation factors where found empirically to obtain a realistic lesion appearance in the simulated dMRI.

\section{Supplementary results}
\subsection{In-vivo denoising}
\begin{figure}[ht!]
\centering
\subfloat{\includegraphics[width=0.5\textwidth]{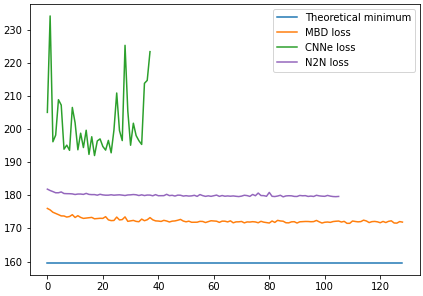}}\hfill
\caption{Validation loss values (mean squared error) obtained by different neural networks and their test errors. Clinical brain dMRI. MBD achieved the lowest loss value}
\label{ckdeval}
\end{figure}

\begin{figure}[ht!]
\centering
\subfloat{\includegraphics[width=0.5\textwidth]{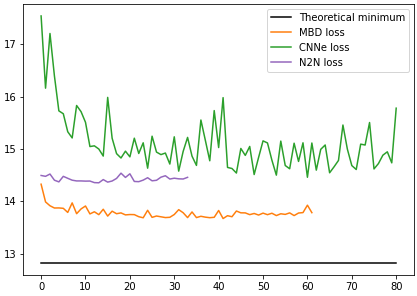}}\hfill
\caption{Validation loss values (mean squared error) obtained by different neural networks and their test errors. STE brain dMRI. MBD achieved the lowest loss value}
\label{steeval}
\end{figure}

Figures~\ref{ckdeval} and \ref{steeval} show validation curves of N2N, CNNe and MBD neural networks. MBD achieves the lowest validation loss in both clinical and STE dMRI denoising training. The validation curve for CNNe shows an unstable decreasal, which suggests it is troublesome to find the optimal solution. Flat lines mark the theoretical loss minimum computed as $MSE = \frac{Var[N_{I_1}]+Var[N_{I_2}]}{2}=\frac{Var[I_1-I_2]}{2}$, where $I_1$ and $I_2$ are output and target images, respectively, and $N$ are noise maps, unobserved directly, but extracted using the difference of  $I_1$ and $I_2$. MBD still has denoising error, equal to the distance between the validation loss value and the theoretical minimum.

\subsection{Lesion denoising in a phantom}

\begin{figure*}[ht!]
\centering
\subfloat[Lesions]{\includegraphics[width=0.5\textwidth]{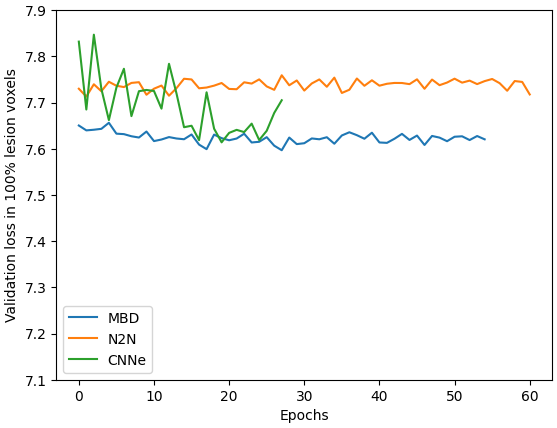}}\hfill
\subfloat[Healthy tissue]{\includegraphics[width=0.5\textwidth]{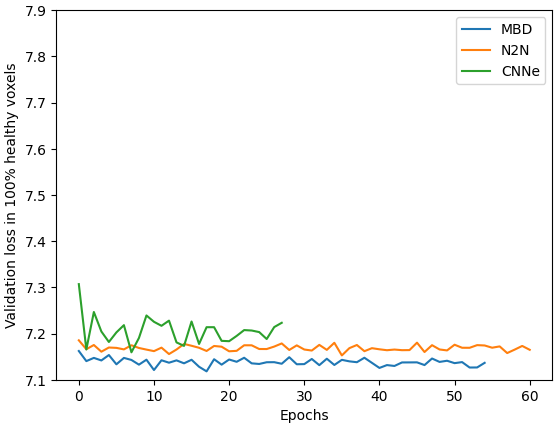}}\hfill
\caption{Validation curves for MBD, N2N and CNNe tracked for lesion and healthy voxels. Only voxels with 100\% content of lesion or healthy tissue are considered. MBD yields lower loss and more stable training}
\label{f:lesioncurves}
\end{figure*}

\begin{figure*}[ht!]
\centering
\subfloat{\includegraphics[width=0.38\columnwidth]{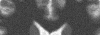}}\hfill
\subfloat{\includegraphics[width=0.38\columnwidth]{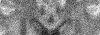}}\hfill
\subfloat{\includegraphics[width=0.38\columnwidth]{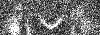}}\hfill
%
%

\subfloat{\includegraphics[width=0.38\columnwidth]{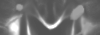}}\hfill

\subfloat{\includegraphics[width=0.38\columnwidth]{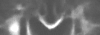}}\hfill
\subfloat{\includegraphics[width=0.38\columnwidth]{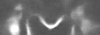}}\hfill
\subfloat{\includegraphics[width=0.38\columnwidth]{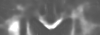}}\hfill
\subfloat{\includegraphics[width=0.38\columnwidth]{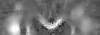}}\hfill
\subfloat{\includegraphics[width=0.38\columnwidth]{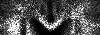}}\hfill

\subfloat{\includegraphics[width=0.38\columnwidth]{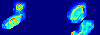}}\hfill
\subfloat{\includegraphics[width=0.38\columnwidth]{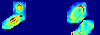}}\hfill
\subfloat{\includegraphics[width=0.38\columnwidth]{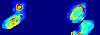}}\hfill
\subfloat{\includegraphics[width=0.38\columnwidth]{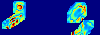}}\hfill
\subfloat{\includegraphics[width=0.38\columnwidth]{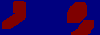}}\hfill

\subfloat{\includegraphics[width=0.38\columnwidth]{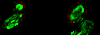}}\hspace{0.01\columnwidth}
\subfloat{\includegraphics[width=0.38\columnwidth]{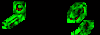}}\hspace{0.01\columnwidth}
\subfloat{\includegraphics[width=0.38\columnwidth]{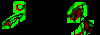}}\hspace{0.01\columnwidth}
\subfloat{\includegraphics[width=0.38\columnwidth]{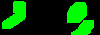}}

\caption{Denoising of the synthetic dMRI with lesions, $f=0.65$, $D_1=400\frac{mm^2}{s}$, $D_2=700 \frac{mm^2}{s}$, $\Delta T_2=0$ ms. The top row shows the b0, b1000, b4000 sequence of noisy DWI. Second top row shows the clean reference image. The three lower rows show denoising results (from left: MBD, N2N, CNNe, MPPCA, ALGe), error maps masked for lesions (from left: MBD, N2N, CNNe, MPPCA, ALGe), maps of error difference (from left: MBD vs. CNNe, MBD vs. N2N, MBD vs. MPPCA, MBD vs. ALGe). Shades of green in the error difference maps mean MBD had lower error, shades of red mean MBD had higher error. MBD shows better denoising for most lesion voxels as compared to other methods}
\label{f:lesionset2}
\end{figure*}

\begin{figure*}[ht!]
\centering
\subfloat{\includegraphics[width=0.38\columnwidth]{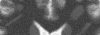}}\hfill
\subfloat{\includegraphics[width=0.38\columnwidth]{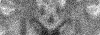}}\hfill
\subfloat{\includegraphics[width=0.38\columnwidth]{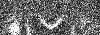}}\hfill
%
%

\subfloat{\includegraphics[width=0.38\columnwidth]{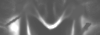}}\hfill

\subfloat{\includegraphics[width=0.38\columnwidth]{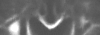}}\hfill
\subfloat{\includegraphics[width=0.38\columnwidth]{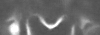}}\hfill
\subfloat{\includegraphics[width=0.38\columnwidth]{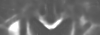}}\hfill
\subfloat{\includegraphics[width=0.38\columnwidth]{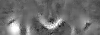}}\hfill
\subfloat{\includegraphics[width=0.38\columnwidth]{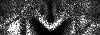}}\hfill

\subfloat{\includegraphics[width=0.38\columnwidth]{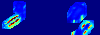}}\hfill
\subfloat{\includegraphics[width=0.38\columnwidth]{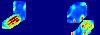}}\hfill
\subfloat{\includegraphics[width=0.38\columnwidth]{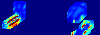}}\hfill
\subfloat{\includegraphics[width=0.38\columnwidth]{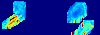}}\hfill
\subfloat{\includegraphics[width=0.38\columnwidth]{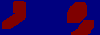}}\hfill

\subfloat{\includegraphics[width=0.38\columnwidth]{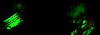}}\hspace{0.01\columnwidth}
\subfloat{\includegraphics[width=0.38\columnwidth]{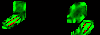}}\hspace{0.01\columnwidth}
\subfloat{\includegraphics[width=0.38\columnwidth]{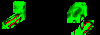}}\hspace{0.01\columnwidth}
\subfloat{\includegraphics[width=0.38\columnwidth]{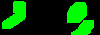}}

\caption{Denoising of the synthetic dMRI with lesions, $f=0.65$, $D_1=700\frac{mm^2}{s}$, $D_2=1100 \frac{mm^2}{s}$, $\Delta T_2=20$ ms. The top row shows the b0, b1000, b4000 sequence of noisy DWI. Second top row shows the clean reference image. The three lower rows show denoising results (from left: MBD, N2N, CNNe, MPPCA, ALGe), error maps masked for lesions (from left: MBD, N2N, CNNe, MPPCA, ALGe), maps of error difference (from left: MBD vs. CNNe, MBD vs. N2N, MBD vs. MPPCA, MBD vs. ALGe). Shades of green in the error difference maps mean MBD had lower error, shades of red mean MBD had higher error. MBD shows better denoising for most lesion voxels as compared to other methods}
\label{f:lesionset4}
\end{figure*}

\begin{figure*}[ht!]
\centering
\subfloat{\includegraphics[width=0.38\columnwidth]{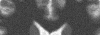}}\hfill
\subfloat{\includegraphics[width=0.38\columnwidth]{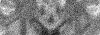}}\hfill
\subfloat{\includegraphics[width=0.38\columnwidth]{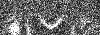}}\hfill
%
%

\subfloat{\includegraphics[width=0.38\columnwidth]{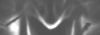}}\hfill

\subfloat{\includegraphics[width=0.38\columnwidth]{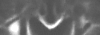}}\hfill
\subfloat{\includegraphics[width=0.38\columnwidth]{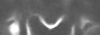}}\hfill
\subfloat{\includegraphics[width=0.38\columnwidth]{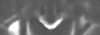}}\hfill
\subfloat{\includegraphics[width=0.38\columnwidth]{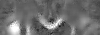}}\hfill
\subfloat{\includegraphics[width=0.38\columnwidth]{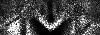}}\hfill

\subfloat{\includegraphics[width=0.38\columnwidth]{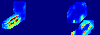}}\hfill
\subfloat{\includegraphics[width=0.38\columnwidth]{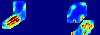}}\hfill
\subfloat{\includegraphics[width=0.38\columnwidth]{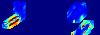}}\hfill
\subfloat{\includegraphics[width=0.38\columnwidth]{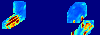}}\hfill
\subfloat{\includegraphics[width=0.38\columnwidth]{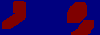}}\hfill

\subfloat{\includegraphics[width=0.38\columnwidth]{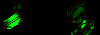}}\hspace{0.01\columnwidth}
\subfloat{\includegraphics[width=0.38\columnwidth]{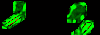}}\hspace{0.01\columnwidth}
\subfloat{\includegraphics[width=0.38\columnwidth]{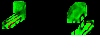}}\hspace{0.01\columnwidth}
\subfloat{\includegraphics[width=0.38\columnwidth]{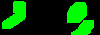}}
\caption{Denoising of the synthetic dMRI with lesions, $f=0.65$, $D_1=700\frac{mm^2}{s}$, $D_2=1100 \frac{mm^2}{s}$, $\Delta T_2=0$ ms. The top row shows the b0, b1000, b4000 sequence of noisy DWI. Second top row shows the clean reference image. The three lower rows show denoising results (from left: MBD, N2N, CNNe, MPPCA, ALGe), error maps masked for lesions (from left: MBD, N2N, CNNe, MPPCA, ALGe), maps of error difference (from left: MBD vs. CNNe, MBD vs. N2N, MBD vs. MPPCA, MBD vs. ALGe). Shades of green in the error difference maps mean MBD had lower error, shades of red mean MBD had higher error. MBD shows better denoising for most lesion voxels as compared to other methods}
\label{f:lesionset5}
\end{figure*}

\begin{figure*}[ht!]
\centering
\subfloat{\includegraphics[width=0.38\columnwidth]{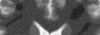}}\hfill
\subfloat{\includegraphics[width=0.38\columnwidth]{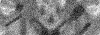}}\hfill
\subfloat{\includegraphics[width=0.38\columnwidth]{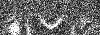}}\hfill
%
%

\subfloat{\includegraphics[width=0.38\columnwidth]{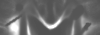}}\hfill

\subfloat{\includegraphics[width=0.38\columnwidth]{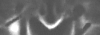}}\hfill
\subfloat{\includegraphics[width=0.38\columnwidth]{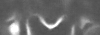}}\hfill
\subfloat{\includegraphics[width=0.38\columnwidth]{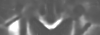}}\hfill
\subfloat{\includegraphics[width=0.38\columnwidth]{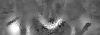}}\hfill
\subfloat{\includegraphics[width=0.38\columnwidth]{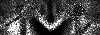}}\hfill

\subfloat{\includegraphics[width=0.38\columnwidth]{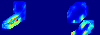}}\hfill
\subfloat{\includegraphics[width=0.38\columnwidth]{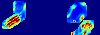}}\hfill
\subfloat{\includegraphics[width=0.38\columnwidth]{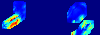}}\hfill
\subfloat{\includegraphics[width=0.38\columnwidth]{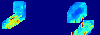}}\hfill
\subfloat{\includegraphics[width=0.38\columnwidth]{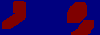}}\hfill

\subfloat{\includegraphics[width=0.38\columnwidth]{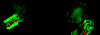}}\hspace{0.01\columnwidth}
\subfloat{\includegraphics[width=0.38\columnwidth]{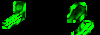}}\hspace{0.01\columnwidth}
\subfloat{\includegraphics[width=0.38\columnwidth]{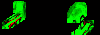}}\hspace{0.01\columnwidth}
\subfloat{\includegraphics[width=0.38\columnwidth]{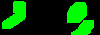}}

\caption{Denoising of the synthetic dMRI with lesions, $f=0.65$, $D_1=700\frac{mm^2}{s}$, $D_2=1100 \frac{mm^2}{s}$, $\Delta T_2=-20$ ms. The top row shows the b0, b1000, b4000 sequence of noisy DWI. Second top row shows the clean reference image. The three lower rows show denoising results (from left: MBD, N2N, CNNe, MPPCA, ALGe), error maps masked for lesions (from left: MBD, N2N, CNNe, MPPCA, ALGe), maps of error difference (from left: MBD vs. CNNe, MBD vs. N2N, MBD vs. MPPCA, MBD vs. ALGe). Shades of green in the error difference maps mean MBD had lower error, shades of red mean MBD had higher error. MBD shows better denoising for most lesion voxels as compared to other methods}
\label{f:lesionset6}
\end{figure*}

Errors of lesion denoising were studied for chosen parameter sets, including lesions that were hypo-, iso- and hyperintense in b0 and lesions that appeared hypo-, iso- or hyperintense in b4000. Figures~\ref{f:lesionset2}-\ref{f:lesionset6} show the test slice, cropped to the nearly frontal brain area, which shows four lesions: a medium-sized round, two elongated with medium width and a big oval. These shapes correspond to lesions No 2, 4, 7 and 8 in Figure~\ref{f:lesion4}. Error maps were masked to focus on regions with a non-zero content of lesion tissue. To avoid random errors resulting from specific noise instances, the computed error maps are an average of 100 repetitions. To enable a clearer comparison of error maps, we computed error difference maps for MBD-CNNe, MBD-N2N, MBD-MPPCA and MBD-ALGe pairs. These maps are shown in pseudocolour in the lower rows of Figures~\ref{f:lesionset2}-\ref{f:lesionset6}, with the green colour denoting superiority of MBD, and red denoting superiority of the other method.

Cases presented here support the finding of the main paper, that MBD was superior to other methods in lesion denoising. However, it is visible that denoising peformance and the resolution preservation is not constant accross different lesion parameter sets. 

